\documentclass[10pt,journal,final,twoside,a4paper]{IEEEtran}
\usepackage[dvipsnames]{xcolor}
\usepackage[pdftex]{graphicx}
\usepackage[utf8]{inputenc}

\usepackage{amsmath}
\usepackage{amsthm}
\usepackage{amsfonts}
\usepackage{array,multirow}
\usepackage{units}
\usepackage[amssymb]{SIunits}
\usepackage{upgreek}
\usepackage{todonotes}
\usepackage{algorithm,algorithmic}
\usepackage{cite}
\usepackage{mathtools}

\usepackage[acronym]{glossaries}

\newcommand{\expc}[1]{\ensuremath{\mathrm{E}\left\{#1\right\}}}

\newcommand{\R}{\mathbb{R}} 
 
\newcommand{\N}{\mathbb{N}}

\newcommand{\veksym}[1]{\ensuremath{\boldsymbol{#1}}}    
\newcommand{\mat}[1]{\ensuremath{\boldsymbol{#1}}}    

\DeclareMathAlphabet\mathboldsf{OT1}{cmss}{bx}{n}

\newcommand{\T}{\mathsf{T}}

\newcommand{\rund}[1]{\left(#1\right)}

\newcommand{\curly}[1]{\left\{#1\right\}}

\newcommand{\dis}{\mathrm{d}}
\newcommand{\lev}{\mathrm{lev}}
\newcommand{\inprod}[1]{\left\langle#1\right\rangle}

\newcommand{\kernel}{\kappa}

\newcommand{\idxIter}{k}
\newcommand{\idxNode}{j}
\newcommand{\idxNeigh}{i}
\newcommand{\idxNodeNeigh}{\idxNode\idxNeigh}

\newcommand{\idxDict}{\ell}

\newcommand{\funTrue}{\psi}
\newcommand{\funEst}{\varphi}
\newcommand{\funEstNode}{\funEst_{\idxNode}}

\newcommand{\funCost}{\Theta}

\newcommand{\funCostNodeTime}{\funCost_{\idxNode,\idxIter}}

\newcommand{\stepSize}{\ensuremath{\mu}}
\newcommand{\stdev}{\ensuremath{\sigma}}
\newcommand{\variance}{\stdev^2}
\newcommand{\kernelBW}{\ensuremath{\zeta}}

\newcommand{\scaOutput}{d}

\newcommand{\scaOutputNodeTime}{\scaOutput_{\idxNode,\idxIter}}

\newcommand{\scaInput}{x}
\newcommand{\scaNoise}{n}

\newcommand{\scaNoiseNodeTime}{\scaNoise_{\idxNode,\idxIter}}

\newcommand{\scaWeight}{w}

\newcommand{\noiseVar}{\variance_{\scaNoise}} 

\newcommand{\vecc}[1]{\veksym{#1}}

\newcommand{\vecInput}{\veksym{\scaInput}}
\newcommand{\vecInputNode}{\vecInput_{\idxNode}}

\newcommand{\vecInputNodeTime}{\vecInput_{\idxNode,\idxIter}}

\newcommand{\vecDict}{\bar{\veksym{\scaInput}}}
\newcommand{\vecDictIdx}{\bar{\veksym{\scaInput}}_{\idxDict}}

\newcommand{\vecWeight}{\veksym{\scaWeight}}
\newcommand{\vecWeightNode}{\vecWeight_{\idxNode}}
\newcommand{\vecWeightNodeTime}{\vecWeight_{\idxNode,\idxIter}}

\newcommand{\vecWeightNeigh}{\vecWeight_{\idxNeigh}}
\newcommand{\vecWeightNeighTime}{\vecWeight_{\idxNeigh,\idxIter}}
\newcommand{\vecWeightOpt}{\veksym{\scaWeight}^{\star}}

\newcommand{\vecKernel}{\veksym{\kernel}}

\newcommand{\matKernel}{\veksym{K}}
\newcommand{\matKernelFull}{\veksym{\mathcal{K}}}

\newcommand{\matEye}{\veksym{I}}
\newcommand{\matConsensus}{\veksym{P}}
\newcommand{\matEdges}{\veksym{G}}

\newcommand{\noNodes}{J}
\newcommand{\noDict}{r}
\newcommand{\noKernels}{Q}

\newcommand{\setNeighNode}{\mathcal{N}_{\idxNode}}
\newcommand{\setNodes}{\mathcal{J}}
\newcommand{\setEdges}{\mathcal{E}}
\newcommand{\setGraph}{\mathcal{G}}
\newcommand{\setDict}{\mathcal{D}}

\newcommand{\setKernels}{\mathcal{Q}}
\newcommand{\setSolution}{\Upsilon}
\newcommand{\setSolutionOpt}{\setSolution^\star}
\newcommand{\setSolutionTime}{\setSolution_\idxIter}

\newcommand{\hslab}{\mathcal{S}}

\newcommand{\hslabNodeTime}{\hslab_{\idxNode,\idxIter}}

\newcommand{\spaceRKHS}{\mathcal{H}}
\newcommand{\spaceDict}{\mathcal{M}}

\newcommand{\spaceInput}{\mathcal{X}}
\newcommand{\spaceOutput}{\mathbb{R}}
\newcommand{\spaceProd}{\mathcal{H}^\times}
\newcommand{\spaceSum}{\mathcal{H}^+}

\newacronym{lms}{LMS}{least-mean-squares}
\newacronym{klms}{KLMS}{kernel least-mean-squares}
\newacronym{knlms}{KNLMS}{kernel normalized least-mean-squares}
\newacronym{mklms}{MKLMS}{multikernel least-mean-squares}
\newacronym{apa}{APA}{affine projection algorithm}
\newacronym{kapa}{KAPA}{kernel affine projection algorithm}
\newacronym{rls}{RLS}{recursive least squares}
\newacronym{krls}{KRLS}{kernel recursive least squares}
\newacronym{mmse}{MMSE}{minimum mean square error}
\newacronym{rkhs}{RKHS}{reproducing kernel Hilbert space}
\newacronym{wsn}{WSN}{wireless sensor network}
\newacronym{wsns}{WSNs}{wireless sensor networks}
\newacronym{dice}{DiCE}{distributed consensus-based estimation}
\newacronym{admm}{ADMM}{alternating direction method of multipliers}
\newacronym{fc}{FC}{fusion center}
\newacronym{ls}{LS}{least squares}
\newacronym{kdice}{KDiCE}{kernel distributed consensus-based estimation}
\newacronym{mkdice}{MKDiCE}{multikernel distributed consensus-based estimation}
\newacronym{mse}{MSE}{mean square error}
\newacronym{kls}{KLS}{kernel least squares}
\newacronym{wrt}{w.r.t.}{with respect to}
\newacronym{sn}{SN}{sensor network}
\newacronym{sns}{SNs}{sensor networks}
\newacronym{cvt}{CVT}{centroidal Voronoi tessellation}
\newacronym{svr}{SVR}{support vector regression}
\newacronym{inp}{INP}{in-network processing}
\newacronym{apsm}{APSM}{adaptive projected subgradient method}
\newacronym{dchypass}{D-CHYPASS}{diffusion-based CHYPASS}
\newacronym{chypass}{CHYPASS}{Cartesian HYPASS}
\newacronym{hypass}{HYPASS}{hyperplane projection along affine subspace}
\newacronym{dmklms}{DMKLMS}{diffusion-based multikernel least-mean-squares}
\newacronym{fatc-klms}{FATC-KLMS}{functional adapt-then-combine KLMS}
\newacronym{norma}{NORMA}{naive online regularized risk minimization algorithm}
\newacronym{rffdklms}{RFF-DKLMS}{random Fourier features diffusion KLMS}

\newlength\figureheight
\newlength\figurewidth

\makeatletter
\newcommand{\thickhline}{%
    \noalign {\ifnum 0=`}\fi \hrule height 1.5pt
    \futurelet \reserved@a \@xhline
}
\newcolumntype{"}{@{\hskip\tabcolsep\vrule width 1.5pt\hskip\tabcolsep}}
\makeatother

\makeatletter
\renewcommand{\todo}[2][]{\tikzexternaldisable\@todo[#1]{#2}\tikzexternalenable}
\makeatother

\newtheorem{definition}{Definition}
\newtheorem{theorem}{Theorem} 
\newtheorem{lemma}{Lemma}
\newtheorem{remark}{Remark}
\newtheorem{fact}{Fact}

\begin{document}
%
\title{Distributed Adaptive Learning with Multiple Kernels in Diffusion Networks}
%
%
%



\author{Ban-Sok~Shin, \IEEEmembership{Student Member, IEEE,} Masahiro~Yukawa, \IEEEmembership{Member, IEEE}, Renato~L.~G.~Cavalcante, \IEEEmembership{Member, IEEE}, and Armin Dekorsy, \IEEEmembership{Senior Member, IEEE}
\thanks{B.-S. Shin and A. Dekorsy are with the Department of Communications Engineering, University of Bremen, Germany, e-mails: shin@ant.uni-bremen.de, dekorsy@ant.uni-bremen.de.}
\thanks{M. Yukawa is with the Department of Electronics and Electrical Engineering, Keio University, Yokohama, Japan, e-mail: yukawa@elec.keio.ac.jp.}
\thanks{R.L.G. Cavalcante is with the Fraunhofer Heinrich Hertz Institute, Berlin, Germany, e-mail: renato.cavalcante@hhi.fraunhofer.de.}
\thanks{M. Yukawa is thankful to JSPS Grants-in-Aid (15K06081, 15K13986, 15H02757).}
}

%
%

\maketitle

\begin{abstract}
We propose an adaptive scheme for distributed learning of nonlinear functions by a network of nodes. The proposed algorithm consists of a local adaptation stage utilizing multiple kernels with projections onto hyperslabs and a diffusion stage to achieve consensus on the estimates over the whole network. Multiple kernels are incorporated to enhance the approximation of functions with several high and low frequency components common in practical scenarios. We provide a thorough convergence analysis of the proposed scheme based on the metric of the Cartesian product of multiple reproducing kernel Hilbert spaces. To this end, we introduce a modified consensus matrix considering this specific metric and prove its equivalence to the ordinary consensus matrix. Besides, the use of hyperslabs enables a significant reduction of the computational demand with only a minor loss in the performance. Numerical evaluations with synthetic and real data are conducted showing the efficacy of the proposed algorithm compared to the state of the art schemes.
%
\end{abstract}

\glsresetall 

\section{Introduction}
\subsection{Background}
Distributed learning within networks is a topic of high importance due to its applicability in various areas such as environmental monitoring, social networks and big data \cite{Ilyas2013,Ingelrest2010,Facchinei2015}. In such applications, observed data are usually spread over the nodes, and thus, they are unavailable at a central entity. In environmental monitoring applications, for instance, nodes observe a common physical quantity of interest such as temperature, gas or humidity at each specific location. For a spatial reconstruction of the physical quantity over the area covered by the network non-cooperative strategies will not deliver a satisfactory performance. Rather distributed learning algorithms relying on information exchanges among neighboring nodes are required to fully exploit the observations available in the network. 


Distributed learning of linear functions has been addressed by a variety of algorithms in the past decade, e.g. \cite{Sayed2014,Paul2013,Mateos2012,Tu2011,Cavalcante2011,Cattivelli2010,Cavalcante2009,Zhu2010,Schizas2009}. In contrast to these works, we address the problem of distributed learning of \emph{nonlinear functions/systems}. To this end, we exploit kernel methods, which have been used to solve e.g. nonlinear regression tasks \cite{Scholkopf2001,Hofmann2008}. Based on a problem formulation in a \gls{rkhs} linear techniques can be applied to approximate an unknown nonlinear function. This function is then modeled as an element of the \gls{rkhs}, and corresponding kernel functions are utilized for its approximation. 
This methodology has been exploited to derive a variety of kernel adaptive filters \cite{Kivinen2004,Engel2004a,VanVaerenbergh2006,Liu2008,Liu2009,Richard2009,Liu2010,VanVaerenbergh2012,Yukawa2012b,Gao2013,Gao2014a}. In particular, the naive online regularized risk minimization \cite{Kivinen2004}, the kernel normalized least-mean-squares, the kernel affine projection \cite{Richard2009} or the \gls{hypass} \cite{Yukawa2012b,Takizawa2015} enjoy significant attention due to their limited complexity and their applicability in online learning scenarios. 
The \gls{hypass} algorithm has been derived from a functional space approach based on the \gls{apsm} \cite{Yamada2004} in the set-theoretic estimation framework \cite{Combettes1993,Theodoridis2011}. It exploits a metric with regard to the kernel Gram matrix showing faster convergence and improved steady-state performance. 
The kernel Gram matrix determines the metric of an \gls{rkhs} and is decisive for the convergence behavior of gradient-descent algorithms \cite{Yukawa2016}.
In \cite{Yukawa2012,Yukawa2015} kernel adaptive filters have been extended by multiple kernels to increase the degree of freedom in the estimation process. By this, a more accurate approximation of functions with several high and low frequency components is possible with a smaller number of dictionary samples compared to using a single kernel only.

Regarding distributed kernel-based estimation algorithms, several schemes have been derived \cite{Shin2016,Shin2017a,Gao2015,Bouboulis2018,Chouvardas2016,Honeine2009a,Honeine2010,Predd2006,Forero2010,Shin2018}. In \cite{Shin2016} a distributed consensus-based regression algorithm based on kernel least squares has been proposed and extended by multiple kernels in \cite{Shin2017a}. Both schemes utilize \gls{admm} \cite{Boyd2010} for distributed consensus-based processing. 
Recent works in \cite{Gao2015,Chouvardas2016,Bouboulis2018} apply diffusion-based schemes to the \gls{klms} to derive distributed kernel adaptive filters where nodes process information in parallel. The \gls{fatc-klms} proposed in \cite{Gao2015} is a kernelized version of the algorithm derived in \cite{Cattivelli2010}. The \gls{rffdklms} proposed in \cite{Bouboulis2018} uses random Fourier features to achieve a fixed-size coefficient vector and to avoid an a priori design of a dictionary set. However, the achievable performance strongly depends on the number of utilized Fourier features. Besides, the aforementioned schemes incorporate update equations in the ordinary Euclidean space and thus, do not exploit the metric induced by the kernel Gram matrix. Furthermore, the majority of these schemes do not consider multiple kernels in their adaptation mechanism.

\subsection{Main Contributions}
\textcolor{black}{
For the derivation of the proposed algorithm we rely on the previous work of \cite{Cavalcante2009}. However, while \cite{Cavalcante2009} only considers distributed learning for linear functions in a Euclidean space we specifically derive a kernel-based learning scheme in an \gls{rkhs} and its isomorphic Euclidean space, respectively. More specifically, we propose a distributed algorithm completely operating in the Cartesian product space of multiple \gls{rkhs}s. The Cartesian product space has been exploited by the \gls{chypass} algorithm for adaptive learning with multiple kernels proposed in \cite{Yukawa2015}. When operating in the corresponding Euclidean parameter space a metric based on the kernel Gram matrix of each employed kernel needs to be considered. This metric is determined by a block diagonal matrix of which the block diagonals are given by kernel Gram matrices. To derive a distributed learning scheme we rely on average consensus on the coefficient vectors for each kernel.
The key idea of our proposed scheme is to fully conduct distributed learning in a Euclidean space considering the metric of the Cartesian product space. This metric is responsible for an enhanced convergence speed of the adaptive algorithm. Operating with this metric implies that the consensus matrix used for diffusion of information within the network needs to be adapted to it. To this end, we introduce a \emph{modified consensus matrix} operating in the metric of the product space. In fact, we show that the modified consensus matrix coincides with the consensus matrix operating in the ordinary Euclidean space as used in \cite{Cavalcante2009}. This finding actually implies that the metric of the product space does not alter the convergence properties of the average consensus scheme. This is particularly important in proving the monotone approximation property of our proposed scheme. We provide a thorough convergence analysis considering the metric of the product space. Specifically, we prove monotone approximation, asymptotic optimization, asymptotic consensus, convergence and characterization of the limit point within the framework of \gls{apsm}. 
As a practical implication we demonstrate that by projecting the current estimate onto a \emph{hyperslab} instead of the ordinary hyperplane we can significantly reduce the computational demand per node. By varying the hyperslab thickness (similar to an error bound), a trade-off between error performance and complexity per node can be adjusted.
We corroborate our findings by extensive numerical evaluations on synthetic as well as real data and by mathematical proofs given in the appendices.}

\section{Preliminaries}
\subsection{Basic Definitions}
\textcolor{black}{
We denote the inner product and the norm of the Euclidean space $\R^M$ by $\langle \cdot,\cdot \rangle_{\R^M}$ and $||\cdot||_{\R^M}$, respectively, and those in the \gls{rkhs} $\spaceRKHS$ by $\langle \cdot,\cdot \rangle_{\spaceRKHS}$ and $||\cdot||_{\spaceRKHS}$, respectively. Given a positive definite matrix $\mat{K}\in\R^{M\times M}$, $\langle \vecc{x},\vecc{y} \rangle_{\mat{K}}:= \vecc{x}^\T\mat{K}\vecc{y}$, $\vecc{x},\vecc{y}\in\R^M$, defines an inner product with the norm $||\vecc{x}||_{\mat{K}}:=\sqrt{\langle \vecc{x},\vecc{x} \rangle_{\mat{K}}}$.
The norm of a matrix $\mat{X}\in\R^{M\times M}$ induced by the vector norm $||\cdot||_{\mat{K}}$ is defined as $||\mat{X}||_{\mat{K}} := \max_{\vecc{y}\neq\vecc{0}} ||\mat{X}\vecc{y}||_{\mat{K}}/||\vecc{y}||_{\mat{K}}$.
The spectral norm of a matrix is denoted as $||\mat{X}||_2$ when we choose $\mat{K}=\matEye_M$ as the $M\times M$ identity matrix \cite{Horn2013}.
A set $C\subset\R^M$ is said to be convex if $\alpha\vecc{x}+(1-\alpha)\vecc{y}\in C$, $\forall\vecc{x},\vecc{y}\in C$, $\forall \alpha\in(0,1)$. If in addition the set $C$ is closed, we call it a closed convex set. The $\matKernel$-projection of a vector $\vecc{w}\in\R^M$ onto a closed convex set $C$ is defined by \cite{Stark1998,Luenberger1969}}
\begin{equation}
	P^{\mat{K}}_{C}(\vecc{w}):=\min_{\vecc{v}\in C} ||\vecc{w}-\vecc{v}||_{\mat{K}}.
\end{equation}
%


\subsection{Multikernel Adaptive Filter}
In the following we present the basics regarding multikernel adaptive filters which have been applied to online regression of nonlinear functions \cite{Yukawa2012,Yukawa2015}. We denote a multikernel adaptive filter by $\funEst:\spaceInput\to\spaceOutput$ where $\spaceInput\subseteq\R^{L}$ is the input space of dimension $L$ and $\R$ the output space. The filter/function $\funEst$ employs $\noKernels$ positive definite kernels $\kernel_q:\spaceInput\times\spaceInput\to\R$ with $q\in\setKernels=\{1,2,\ldots,Q\}$. Each kernel $\kernel_q$ induces an \gls{rkhs} $\spaceRKHS_q$ \cite{Scholkopf2001}, and $\funEst$ uses corresponding dictionaries $\setDict_q=\{\kernel_q(\cdot,\vecDictIdx)\}_{\idxDict=1}^{\noDict}$, each of cardinality $\noDict$. Here, each dictionary $\setDict_q$ contains kernel functions $\kernel_q$ centered at samples $\vecDict_{\idxDict}\in\spaceInput$. For simplicity, we assume that each dictionary $\setDict_q$ uses the same centers $\{ \vecDictIdx \}_{\idxDict=1}^\noDict$ although this assumption is not required. The multikernel adaptive filter $\funEst$ is then given by
\begin{equation}
	\funEst := \sum_{q\in\setKernels} \sum_{\idxDict=1}^{\noDict} \scaWeight_{q,\idxDict} \kernel_q(\cdot,\vecDictIdx). \label{eq:funMkaf}
\end{equation}
%
The output of $\funEst$ for arbitrary input samples $\vecInput$ can be computed via
\begin{equation}
	\funEst(\vecInput) 
					   = \sum_{q\in\setKernels} \sum_{\idxDict=1}^{\noDict} \scaWeight_{q,\idxDict} \kernel_q(\vecInput,\vecDictIdx)
					   = \left\langle \vecWeight,\vecKernel(\vecInput) \right\rangle_{\R^{\noDict\noKernels}}. \label{eq:estimatedOutputEuclidean}
\end{equation}
Here, vectors $\vecWeight$ and $\vecKernel(\vecInput)$ are defined as
\begin{align*}
	\vecWeight_{(q)} &:= [\scaWeight_{q,1},\ldots,\scaWeight_{q,\noDict}]^\T \in\R^{\noDict}, \\
	\vecWeight &:=[\vecWeight_{(1)}^\T,\ldots,\vecWeight_{(Q)}^\T]^\T \in\R^{\noDict\noKernels},\\
	\vecKernel_{q}(\vecInput) &:= [\kernel_q(\vecInput,\vecDict_1),\ldots,\kernel_q(\vecInput,\vecDict_\noDict)]^\T \in\R^{\noDict}, \\
	\vecKernel(\vecInput) &:=[\vecKernel_{1}^\T(\vecInput),\ldots,\vecKernel_{Q}^\T(\vecInput)]^\T \in\R^{\noDict\noKernels}.
\end{align*}
A commonly used kernel function is the Gaussian kernel defined as 
\begin{equation}
	\kernel_q(\vecInput_1,\vecInput_2) = \exp\rund{-\frac{||\vecInput_1-\vecInput_2||_{\R^L}^2}{2\kernelBW_q^2}},\, \vecInput_1,\vecInput_2\in\spaceInput,
	\label{eq:gausskernel}
\end{equation}
where $\kernelBW_q>0$ is the kernel bandwidth. The metric of an \gls{rkhs} is determined by the kernel Gram matrix. It contains the inherent correlations of a dictionary $\setDict_q$ \gls{wrt} the kernel $\kernel_q$ and is defined as
\begin{equation}
	\matKernel_q := \left[\begin{array}{ccc}
				\kernel_q(\vecDict_1,\vecDict_1) & \ldots & \kernel_q(\vecDict_1,\vecDict_\noDict) \\
				\vdots & \ddots & \vdots \\
				\kernel_q(\vecDict_\noDict,\vecDict_1) & \ldots & \kernel_q(\vecDict_\noDict,\vecDict_\noDict)
				\end{array}\right] \in \R^{\noDict\times\noDict}.
\end{equation}
Assuming that each dictionary $\setDict_q$ is linearly independent it follows that each $\matKernel_q$ is positive-definite \cite{Luenberger1969}.
Moreover, we introduce the multikernel Gram matrix $\matKernel~:=~\mathrm{blkdiag}\{\matKernel_1,\matKernel_2,\ldots,\matKernel_Q\}\in \R^{\noDict\noKernels\times\noDict\noKernels}$ being the block-diagonal matrix of the Gram matrices of all kernels. Then, by virtue of Lemma 1 from \cite{Takizawa2016a} we can parameterize $\funEst$ by $\vecWeight$ in the Euclidean space $\R^{\noDict\noKernels}$ using the $\matKernel$ inner product $\langle \cdot,\cdot \rangle_{\matKernel}$. 
\textcolor{black}{
In fact, the $\matKernel$-metric in the Euclidean space corresponds to the metric of the Cartesian product of multiple \gls{rkhs}s defined as $\spaceProd:=\spaceRKHS_1\times\spaceRKHS_2 \times\ldots\times\spaceRKHS_\noKernels:=\{(f_1,f_2,\ldots,f_\noKernels):f_q\in\spaceRKHS_q,q\in\setKernels\}$ \cite{Yukawa2015}.
}
Indeed, we can express \eqref{eq:estimatedOutputEuclidean} equivalently by
\begin{equation}
	\funEst(\vecInput) = \left\langle \vecWeight,\vecKernel(\vecInput) \right\rangle_{\R^{\noDict\noKernels}} = \langle \vecWeight,\matKernel^{-1}\vecKernel(\vecInput) \rangle_{{\matKernel}}.
\end{equation}
Instead of applying a learning method to the function $\funEst$ in $\rund{\spaceProd,\inprod{\cdot,\cdot}_{\spaceProd}}$ we can directly apply it to the coefficient vector $\vecWeight\in\R^{\noDict\noKernels}$ in $\rund{\R^{\noDict\noKernels},\inprod{\cdot,\cdot}_{\matKernel}}$.
This representation is based on the \emph{parameter space approach} from the kernel adaptive filtering literature with the \emph{functional space approach} as its equivalent counterpart, see \cite[Appendix A]{Yukawa2012}. In the following, we will formulate the distributed learning problem in the parameter space $\rund{\R^{\noDict\noKernels},\inprod{\cdot,\cdot}_{\matKernel}}$ to facilitate an easy understanding. However, we emphasize that this formulation originates from considerations in an isomorphic functional space. The interested reader is referred to Appendix \ref{sec:appendix_funSpace} for a problem formulation in the functional space.

\section{Problem Formulation and Objective}

\subsection{System Model}
We address the problem of distributed adaptive learning of a continuous, nonlinear function $\funTrue:\spaceInput\to\R$ by a network of $\noNodes$ nodes. 
\textcolor{black}{
The function $\funTrue$ is assumed to lie in the sum space of $\noKernels$ \gls{rkhs}s defined as $\spaceSum:=\spaceRKHS_1 + \spaceRKHS_2 + \ldots + \spaceRKHS_\noKernels:=\left\{ \sum_{q\in\setKernels} f_q \,|\, f_q\in\spaceRKHS_q \right\}$.
} 
We label a node by index $\idxNode$ and the time by index $\idxIter$. Each node~$\idxNode$ observes the nonlinear function $\funTrue\in\spaceSum$ by sequentially feeding it with inputs $\vecInputNodeTime\in\R^{L}$. Then each node $\idxNode$ acquires the measurement $\scaOutputNodeTime\in\R$ per time index $\idxIter$ via
\begin{equation}
	\scaOutputNodeTime = \funTrue(\vecInputNodeTime)+\scaNoiseNodeTime,
	\label{eq:systemModel}
\end{equation}
where $\scaNoiseNodeTime\in\R$ is a noise sample.
Based on the nodes' observations, at each time index $k$ we have a set of $J$ acquired input-output samples $\{(\vecInputNodeTime,\scaOutputNodeTime)\}_{\idxNode\in\setNodes}$ available within the network.

To describe the connections among the nodes in the network we employ a graph $\mathcal{G}=(\setNodes,\setEdges)$ with a set of nodes $\setNodes=\{1,\ldots,J\}$ and a set of edges $\setEdges\subseteq\setNodes\times\setNodes$. Each edge in the network represents a connection between two nodes $j$ and $i$ given by $(j,i)\in\setEdges$ where each node $\idxNode$ is connected to itself, i.e., $(j,j)\in\setEdges$. We further assume that the graph is \emph{undirected}, i.e., edges $(j,i)$ and $(i,j)$ are equivalent to each other. The set of neighbors for each node $\idxNode$ is given as $\setNeighNode=\{\idxNeigh\in\setNodes\,|\,(\idxNode,\idxNeigh)\in\setEdges\}$ containing all nodes connected to node $j$ (including node $j$ itself). Furthermore, we consider the graph to be \emph{connected}, i.e., each node can be reached by any other node over multiple hops.
The objective of the nodes is to learn the nonlinear function $\funTrue$ based on the acquired input-output samples $\{(\vecInputNodeTime,\scaOutputNodeTime)\}_{\idxNode\in\setNodes}$ in a distributed fashion. To this end, nodes are able to exchange information with their neighboring nodes to enhance their individual estimate of the unknown function $\funTrue$.

\subsection{Problem Formulation in Parameter Space}
Based on the parametrization of the multikernel adaptive filter $\funEst$ by the coefficient vector $\vecWeight$ we formulate an optimization problem in the parameter space of $\vecWeight$. 
The objective is to find a $\vecWeight$ such that the estimated output $\funEst(\vecInput)=\langle \vecWeight,\matKernel^{-1}\vecKernel(\vecInput) \rangle_{\matKernel}$ is close to the function output $\funTrue(\vecInput)$ for arbitrary input samples $\vecInput\in\spaceInput$. This has to be achieved in a distributed fashion for each node $\idxNode$ in the network based on the acquired data pairs $\{(\vecInputNodeTime,\scaOutputNodeTime)\}_{\idxNode\in\setNodes}$. Thus, we equip each node~$\idxNode$ with a multikernel adaptive filter \eqref{eq:funMkaf} parameterized by its individual coefficient vector $\vecWeightNode$. Furthermore, each node~$\idxNode$ is assumed to rely on the same dictionaries $\setDict_q, q\in\setKernels$, i.e., they are globally known and common to all nodes.
To specify the coefficient vectors which result in an estimate close to the node's measurement, we introduce the closed convex set $\hslabNodeTime$ per node $\idxNode$ and time index $k$:
\begin{equation}
	\hslabNodeTime := \curly{ \vecWeightNode\in\R^{\noDict\noKernels}: |\langle \vecWeightNode, \matKernel^{-1}\vecKernel(\vecInputNodeTime) \rangle_{\matKernel}-\scaOutputNodeTime|\leq\varepsilon_{\idxNode} }, \nonumber
\end{equation}
where $\varepsilon_{\idxNode}\geq0$ is a design parameter. The set $\hslabNodeTime$ is a \emph{hyperslab} containing those vectors $\vecWeightNode$ which provide an estimate $\funEst(\vecInputNodeTime)=\inprod{\vecWeightNode, \matKernel^{-1}\vecKernel(\vecInputNodeTime)}_{\matKernel}$ with a maximum distance of $\varepsilon_j$ to the desired output $\scaOutputNodeTime$ \cite{Yamada2002}. The parameter $\varepsilon_j$ controls the thickness of the hyperslab $\hslabNodeTime$, and is introduced to consider the uncertainty caused by measurement noise $\scaNoiseNodeTime$. The key issue is to find an optimal $\vecWeightNode\in\hslabNodeTime$. To this end, we define a local cost function $\funCostNodeTime$ at time $\idxIter$ per node $\idxNode$ as the metric distance between its coefficient vector $\vecWeightNode$ and the hyperslab $\hslabNodeTime$ in the $\matKernel$-norm sense:
\begin{equation}
	\funCostNodeTime(\vecWeightNode) := ||\vecWeightNode-P_{\hslabNodeTime}^{\matKernel}(\vecWeightNode)||_{\matKernel}. \label{eq:localCost}
\end{equation}
This cost function gives the residual between $\vecWeightNode$ and its $\matKernel$-projection onto $\hslabNodeTime$. Due to the distance metric $\funCostNodeTime(\vecWeightNode)$ is a non-negative, convex function with minimum value
	$\funCost_{\idxNode,k}^{\star}:=\underset{\vecWeightNode}{\min}\,\funCostNodeTime(\vecWeightNode) = 0.$
Then we define the global cost of the network at time~$\idxIter$ to be the sum of all local costs by
\begin{equation}
	\funCost_\idxIter(\vecWeightNode):=\sum_{\idxNode\in\setNodes}\funCostNodeTime(\vecWeightNode)
	\label{eq:globalCost}
\end{equation}
where each individual cost $\funCostNodeTime$ can be time-varying. The objective is to minimize the sequence $(\funCost_\idxIter)_{\idxIter\in\N}$ of global costs \eqref{eq:globalCost} over all nodes in the network where due to convexity of $\funCostNodeTime$ the global cost $\funCost_\idxIter$ is also convex.
Simultaneously, the coefficient vectors $\vecWeightNode$ of all nodes have to converge to the same solution, which guarantees consensus in the network. To this end, we consider the following optimization problem at time $\idxIter$ as in \cite{Cavalcante2009,Cavalcante2011,Cavalcante2013}:
\begin{subequations}\label{eq:opt_problem_parameter}
\begin{align}
	\underset{\curly{\vecWeightNode|\idxNode\in\setNodes}}{\min}& \funCost_\idxIter(\vecWeightNode) := \sum_{\idxNode\in\setNodes} 
	\funCostNodeTime(\vecWeightNode)\\
	\mathrm{s.t. } &\quad \vecWeightNode = \vecWeightNeigh, \quad \forall\idxNeigh\in\setNeighNode. \label{eq:consensus_w}
\end{align}
\end{subequations}
Constraint \eqref{eq:consensus_w} enforces all coefficient vectors to converge to the same solution, i.e., $\vecWeight_1 = \vecWeight_2 = \cdots =\vecWeight_J$ guaranteeing consensus within the network.

\subsection{Optimal Solution Set}
\textcolor{black}{
From the definition \eqref{eq:localCost} of the local cost $\funCostNodeTime$ we directly see that its minimizers are given by points in the  hyperslab $\hslabNodeTime$.
Since each local cost $\funCostNodeTime$ is a metric distance with minimum value zero the minimizers of the global cost $\funCost_\idxIter$ at time~$\idxIter$ are given by the intersection $\Upsilon_\idxIter:=\bigcap_{\idxNode\in\setNodes} \hslabNodeTime$. Points in $\Upsilon_\idxIter$ minimize each local cost $\funCostNodeTime$ and therefore also their sum $\funCost_\idxIter(\vecWeightNode)=\sum_{\idxNode\in\setNodes}\funCostNodeTime(\vecWeightNode)$.
Thus, a point minimizing each local cost $\funCostNodeTime,\forall\idxNode\in\setNodes,$ is also a minimizer of the global cost $\funCost_\idxIter$. To consider arbitrary many time instants~$\idxIter\geq 0$ we can now define the optimal solution set to problem~\eqref{eq:opt_problem_parameter}:
\begin{equation}
	\Upsilon^\star:=\bigcap_{\idxIter\geq 0} \bigcap_{\idxNode\in\setNodes} \hslabNodeTime.
\end{equation}
Points in the set $\Upsilon^\star$ minimize the global cost $\funCost_\idxIter$ for any time instant~$\idxIter$ and at any node~$\idxNode$.
We therefore call a point $\vecWeightOpt\in\Upsilon^\star$ \emph{ideal estimate}.
However, finding $\vecWeightOpt$ is a challenging task particularly under practical considerations. Due to limited memory, for instance, not all measurements can be stored over time at each node. Hence, information about the set $\Upsilon^\star$ is unavailable and thus an ideal estimate $\vecWeightOpt$ cannot be acquired. 
An alternative, feasible task is the minimization of all but finitely many global costs $\funCost_\idxIter$. 
This approach stems from the intuition that a good estimate should minimize as many costs $\funCost_\idxIter$ as possible.
To acquire such an estimate the nodes should agree on a point contained in the set
\begin{equation}
	\Upsilon := \overline{\underset{\idxIter\to\infty}{\lim \inf}\, \Upsilon_\idxIter} = \overline{\bigcup_{\idxIter=0}^\infty \bigcap_{m\geq\idxIter} \Upsilon_m} \supset \Upsilon^\star
	\label{eq:bestSet}
\end{equation}
where the overbar gives the closure of a set. Finding a point in $\Upsilon$ is clearly a less restrictive task than finding one in $\Upsilon^\star$ since all global costs $\funCost_\idxIter$ excluding finitely many ones need to be minimized. Therefore, our proposed algorithm should achieve estimates in the set $\Upsilon$. It has been shown that the \gls{apsm} converges to points in the set~$\Upsilon$ \cite{Yamada2004,Slavakis2006}.
\begin{remark}
For the above considerations we need to assume that $\Upsilon^\star\neq \emptyset$. 
To enable $\Upsilon^\star\neq\emptyset$ the hyperslab threshold $\varepsilon_{\idxNode}$ of $\hslabNodeTime$ should be chosen sufficiently large depending on the noise distribution and its variance. Examples on how to choose $\varepsilon_{\idxNode}$ in noisy environments have been proposed in \cite{Yamada2002}. For impulsive noise occurring finitely many times one can regard the time instant of the final impulse as $\idxIter=0$ to guarantee $\Upsilon^\star\neq \emptyset$. If however impulsive noise occurs infinitely many times on the measurements it is not straightforward to ensure $\Upsilon^\star \neq \emptyset$ and convergence of the \gls{apsm} which will be introduced later on. Nevertheless, whenever the impulsive noise occurs the error signal in the \gls{apsm} will abruptly change. Based on this change those noisy measurements can be detected and discarded in practice so that $\Upsilon^\star \neq \emptyset$ is satisfied.
\end{remark}
}

\section{Proposed Algorithm: Diffusion-Based Multikernel Adaptive Filter}
To solve \eqref{eq:opt_problem_parameter} in a distributed way we employ a two-step scheme consisting of a local adaptation and a diffusion stage which has been commonly used in the literature, see e.g. \cite{Cavalcante2013,Sayed2014,Gao2015}:
\begin{enumerate}
	\item a \emph{local \gls{apsm} update} per node $\idxNode$ on the coefficient vector $\vecWeightNode$ giving an intermediate coefficient vector $\vecWeightNode'$;
	\item a \emph{diffusion stage} to fuse vectors $\vecWeightNeigh'$ from neighboring nodes $\idxNeigh\in\setNeighNode$ to update $\vecWeightNode$.
\end{enumerate}
Step 1) ensures that each local cost $\funCostNodeTime$ is reduced, and, hence the global cost $\funCost_\idxIter$ is reduced as well. Step 2) seeks for a consensus among all coefficient vectors $\{\vecWeightNode\}_{\idxNode\in\setNodes}$ through information exchange among neighboring nodes to satisfy constraint \eqref{eq:consensus_w}. By this exchange each node inherently obtains the property sets from its neighbors which can be exploited to improve the convergence behavior of the learning algorithm.

\subsection{Local APSM Update}
The \gls{apsm} asymptotically minimizes a sequence of non-negative convex (not necessarily differentiable) functions \cite{Yamada2004} and can thus be used to minimize the local cost function $\funCostNodeTime(\vecWeightNode)$ in \eqref{eq:localCost} per node $\idxNode$. For the coefficient vector $\vecWeightNodeTime\in\R^{\noDict\noKernels}$ at node $\idxNode$ and time $\idxIter$ a particular case of the \gls{apsm} update with the $\matKernel$-norm reads
\begin{equation}
	\vecWeight'_{\idxNode,k+1} := 
	\begin{cases}	
		\vecWeightNodeTime -\stepSize_{\idxNode,\idxIter} \dfrac{\funCostNodeTime(\vecWeightNodeTime)-\funCost_{\idxNode,k}^\star}{||\funCost'_{\idxNode,k}(\vecWeightNodeTime)||_{\matKernel}^2}  \funCost'_{\idxNode,k}(\vecWeightNodeTime)\\
		\quad \quad \quad \quad \qquad \text{if } \funCost'_{\idxNode,\idxIter}(\vecWeightNodeTime)\neq \vecc{0}\\
		\vecWeightNodeTime \quad \quad \quad  \quad \text{otherwise}
	\end{cases}
\end{equation}
where $\funCost'_{\idxNode,\idxIter}(\vecWeightNodeTime)$ is a subgradient\footnote{A vector $\funCost'(\vecc{y})\in\R^M$ is a subgradient of a function $\funCost:\R^M\to\R$ at $\vecc{y}\in\R^M$ if $\funCost(\vecc{y})+\langle \vecc{x}-\vecc{y} \rangle \funCost'(\vecc{y}) \leq \funCost(\vecc{x})$ for all $\vecc{x}\in\R^M$.} of $\funCost_{\idxNode,\idxIter}(\vecWeightNodeTime)$ at $\vecWeightNodeTime$. The parameter $\stepSize_{\idxNode,\idxIter}\in(0,2)$ is the step size. Since the learning scheme is to operate with the $\matKernel$-metric it is used for the squared norm in the denominator. A subgradient for \eqref{eq:localCost} is given by \cite{Yamada2004}
\begin{equation}
	\funCost'_{\idxNode,k}(\vecWeightNodeTime) = \frac{\vecWeightNodeTime-P^{\matKernel}_{\hslabNodeTime}(\vecWeightNodeTime)}{||\vecWeightNodeTime-P^{\matKernel}_{\hslabNodeTime}(\vecWeightNodeTime)||_{\matKernel}},
	\quad \text{for } \vecWeightNodeTime\notin\hslabNodeTime.
\end{equation}
This subgradient gives $||\funCost'_{\idxNode,k}(\vecWeightNodeTime)||_{\matKernel}^2=1$ and thus we arrive at the following \gls{apsm} update per node~$\idxNode$:
\begin{equation}
	\vecWeight'_{\idxNode,k+1} := \vecWeightNodeTime-\stepSize_{\idxNode,\idxIter} \rund{ \vecWeightNodeTime-P^{\matKernel}_{\hslabNodeTime}(\vecWeightNodeTime) }.
	\label{eq:localAPSM}
\end{equation}
As we can see, the difference vector $\vecWeightNodeTime-P^{\matKernel}_{\hslabNodeTime}(\vecWeightNodeTime)$ is used to move the coefficient vector $\vecWeightNodeTime$ into the direction of the hyperslab $\hslabNodeTime$ controlled by the step size $\stepSize_{\idxNode,\idxIter}$.
Note that this update solely relies on local information, i.e., no information from neighboring nodes is needed. The projection $P^{\matKernel}_{\hslabNodeTime}(\vecWeightNodeTime)$ is calculated by \cite{Stark1998} 
\begin{align}
	P^{\matKernel}_{\hslabNodeTime}(\vecWeight) =
	\begin{dcases}
	\vecWeight, \quad\quad \text{if } \vecWeight\in \hslabNodeTime\\
	\vecWeight-\dfrac{\vecWeight^\T\vecKernel(\vecInput_{j,k})-\scaOutputNodeTime-\varepsilon_j}{||\matKernel^{-1}\vecKernel(\vecInputNodeTime)||_{\matKernel}^2}\matKernel^{-1}\vecKernel(\vecInputNodeTime),\\  \quad\text{if} \quad \vecWeight^\T\vecKernel(\vecInputNodeTime)>\scaOutputNodeTime+\varepsilon_j \\ 
	\vecWeight-\dfrac{\vecWeight^\T\vecKernel(\vecInputNodeTime)-\scaOutputNodeTime+\varepsilon_j}{||\matKernel^{-1}\vecKernel(\vecInputNodeTime)||_{\matKernel}^2} \matKernel^{-1}\vecKernel(\vecInputNodeTime),\\ \quad \text{if} \quad \vecWeight^\T\vecKernel(\vecInputNodeTime)<\scaOutputNodeTime-\varepsilon_j.
	\end{dcases}
	\label{eq:projection_slab}
\end{align}

\subsection{Diffusion Stage}
To satisfy constraint \eqref{eq:consensus_w} and reach consensus on the coefficient vectors $\vecWeightNode$, each node $\idxNode$ fuses its own vector $\vecWeightNode'$ with those of its neighbors $\{\vecWeightNeigh'\}_{\idxNeigh\in\setNeighNode}$. To this end, we employ a symmetric matrix $\mat{G}\in\R^{\noNodes\times\noNodes}$ assigning weights to the edges in the network. 
The $(j,i)$-entry of $\mat{G}$ is denoted by $g_{ji}$ and gives the weight on the edge between nodes $j$ and $i$. Obviously, if no connection is present among both nodes, the entry will be zero. The fusion step per node $j$ at time $\idxIter$ follows
\begin{equation}
	\vecWeightNodeTime := \sum_{\idxNeigh\in\setNeighNode} g_{\idxNodeNeigh} \vecWeight'_{\idxNeigh,k}. 
	\label{eq:local_fusion}
\end{equation}
To guarantee that all nodes converge to the same coefficient vector, $\mat{G}$ needs to fulfill the following conditions \cite{Cavalcante2009}:
\begin{equation}
	||\mat{G}-(1/\noNodes)\vecc{1}_\noNodes\vecc{1}^\T_\noNodes||_2 < 1, \quad \mat{G} \vecc{1}_\noNodes = \vecc{1}_\noNodes,
\end{equation}
where $\vecc{1}_\noNodes$ is the vector of $\noNodes$ ones.
The first condition guarantees convergence to the average of all states in the network while the second condition keeps the network at a stable state if consensus has been reached. Such matrices have been vastly applied in literature for consensus averaging problems, see e.g. \cite{Xiao2004,Boyd2005,Sayed2014}.
Our proposed algorithm to solve \eqref{eq:opt_problem_parameter} is then given by the following update equations per node $\idxNode$ and time index $k$:
\begin{subequations} \label{eq:dchypass}
\begin{align}
	\vecWeight'_{j,k+1} &:= \vecWeightNodeTime-\stepSize_{\idxNode,\idxIter} \rund{ \vecWeightNodeTime-P^{\matKernel}_{\hslabNodeTime}(\vecWeightNodeTime) } \label{eq:localApsm}\\
	\vecWeight_{j,k+1}  &:= \sum_{\idxNeigh\in\setNeighNode} g_{\idxNodeNeigh} \vecWeight'_{\idxNeigh,k+1} 
	\label{eq:diffStep}
\end{align}
\end{subequations}
where the projection $P^{\matKernel}_{\hslabNodeTime}(\vecWeightNodeTime)$ is given in \eqref{eq:projection_slab}. In each iteration $k$ each node $\idxNode$ performs a local APSM update and transmits its intermediate coefficient vector $\vecWeightNodeTime'$ to its neighbors $\idxNeigh\in\setNeighNode$. After receiving the intermediate coefficient vectors $\vecWeightNeighTime'$ from its neighbors, each node~$\idxNode$ fuses these with its own vector $\vecWeightNodeTime'$ by a weighted average step.

In fact, \eqref{eq:localApsm} comprises the projection in the Cartesian product of $\noKernels$ RKHSs which is used by the \gls{chypass} algorithm \cite{Yukawa2015}. Therefore, we call our proposed scheme \gls{dchypass} being a distributed implementation of \gls{chypass}. 
\begin{remark}
\textcolor{black}{
If the diffusion stage \eqref{eq:diffStep} in \gls{dchypass} is omitted the algorithm reduces to a local adaptation or non-cooperative scheme where each node individually approximates $\funTrue$ based on its node-specific measurement data. However, in this case each node~$\idxNode$ has access to its individual property set $\hslabNodeTime$ only per time instant~$\idxIter$. In contrast, by diffusing the coefficient vectors among neighboring nodes each node~$\idxNode$ inherently obtains information about the property sets $\{\hslab_{\idxNeigh,\idxIter}\}_{\idxNeigh\in\setNeighNode}$ of its neighbors. This can be simply observed when inserting \eqref{eq:localApsm} into \eqref{eq:diffStep}. Therefore, compared to local adaptation \gls{dchypass} will show a faster convergence speed and a lower steady-state error due to a cooperation within the network. Several works have shown the benefit of distributed approaches over non-cooperative strategies in the context of diffusion-based adaptive learning, see \cite{Sayed2014} and references therein. }
\end{remark}

\section{Theoretical Analysis}
\subsection{Consensus Matrix}
To analyze the theoretical properties of the \gls{dchypass} algorithm, let us first introduce the definition of the consensus matrix.
\begin{definition}[Consensus Matrix \cite{Cavalcante2009}] \label{def:consensusMat}
	A consensus matrix $\matConsensus\in\R^{\noDict\noKernels\noNodes\times\noDict\noKernels\noNodes}$ is a square matrix satisfying the following two properties.
\begin{enumerate}
	\item $\matConsensus\vecc{z}=\vecc{z}$ and $\matConsensus^\T\vecc{z}=\vecc{z}$ for any vector $\vecc{z}\in\mathcal{C}:=\curly{ \vecc{1}_\noNodes\otimes\vecc{a}\in\R^{\noDict\noKernels\noNodes}\,|\,\vecc{a}\in\R^{\noDict\noKernels}}$.
	\item The $\noDict\noKernels$ largest singular values of $\matConsensus$ are equal to one and the remaining $\noDict\noKernels\noNodes-\noDict\noKernels$ singular values are strictly less than one.
\end{enumerate}
\end{definition}
We denote by $\otimes$ the Kronecker product.
We can further establish the following properties of the consensus matrix $\matConsensus$:
\begin{lemma}[Properties of Consensus Matrix \cite{Cavalcante2009}] \label{lem:consensusMat}
Let $\vecc{e}_n\in\R^{\noDict\noKernels}$ be a unit vector with its $n$-th entry being one and $\vecc{b}_n= (\vecc{1}_{\noNodes}\otimes\vecc{e}_n)/\sqrt{\noNodes}\in\R^{\noDict\noKernels\noNodes}$. Further, we define the \emph{consensus subspace} $\mathcal{C}:=\mathrm{span}\{\vecc{b}_1,\ldots,\vecc{b}_{\noDict\noKernels} \}$ and the stacked vector of all coefficient vectors in the network $\vecc{z}_{\idxIter} = [\vecWeight_{1,k}^\T,\ldots,\vecWeight_{\noNodes,k}^\T]^\T\in\R^{\noDict\noKernels\noNodes}$. Then, we have the following properties.
\begin{enumerate}
	\item The consensus matrix $\matConsensus$ can be decomposed into $\matConsensus=\mat{B}\mat{B}^\T+\mat{X}$ with $\mat{B}:=[\vecc{b}_1 \ldots \vecc{b}_{\noDict\noKernels}]\in\R^{\noDict\noKernels\noNodes\times\noDict\noKernels}$ and $\mat{X}\in\R^{\noDict\noKernels\noNodes\times\noDict\noKernels\noNodes}$ satisfying $\mat{X}\mat{B}\mat{B}^\T=\mat{B}\mat{B}^\T\mat{X}=\vecc{0}$ and $||\mat{X}||_2<1$.
	\item The nodes have reached consensus at time index $k$ if and only if $(\matEye_{\noDict\noKernels\noNodes}-\mat{B}\mat{B}^\T)\vecc{z}_{\idxIter}=\vecc{0}$, i.e., $\vecc{z}_{\idxIter}\in\mathcal{C}$.
\end{enumerate}
\end{lemma}
A consensus matrix can be constructed by matrix $\matEdges$ as $\matConsensus=\matEdges\otimes\matEye_{\noDict\noKernels}$ where $\matEye_{\noDict\noKernels}$ is the $\noDict\noKernels\times\noDict\noKernels$ identity matrix.
The matrix $\matConsensus$ is then said to be compatible to the graph $\setGraph$ since $\vecc{z}_{\idxIter+1}=\matConsensus\vecc{z}_{\idxIter}$ can be equivalently calculated by $\vecWeight_{\idxNode,\idxIter+1}=\sum_{\idxNeigh\in\setNeighNode} g_{\idxNodeNeigh} \vecWeight_{\idxNeigh,\idxIter}$ (see \eqref{eq:local_fusion}) \cite{Cavalcante2009}.
By definition of the consensus matrix we know that $||\matConsensus||_2=1$ holds. However, for further analysis of the \gls{dchypass} algorithm, we need to know the norm \gls{wrt} matrix $\matKernel$ since \gls{dchypass} operates with the $\matKernel$-metric. Therefore, we introduce a modified consensus matrix $\widehat{\matConsensus}$ satisfying $||\widehat{\matConsensus}||_{\matKernelFull}=1$.
\begin{lemma}[Modified Consensus Matrix]\label{lem:modConsensusMat}
	Suppose that $\matConsensus$ is a consensus matrix defined as in Definition \ref{def:consensusMat}. Let $$\widehat{\matConsensus}:=\matKernelFull^{\nicefrac{-1}{2}}\matConsensus\matKernelFull^{\nicefrac{1}{2}}$$ be the \emph{modified consensus matrix} where $\matKernelFull$ is the block-diagonal matrix with $\noNodes$ copies of $\matKernel$:
	\begin{equation}
		\matKernelFull := \matEye_{\noNodes} \otimes \matKernel \in\R^{\noDict\noKernels\noNodes\times\noDict\noKernels\noNodes}.
	\end{equation}
	Assume further, that the dictionary $\setDict_q=\{ \kernel_q(\cdot,\vecDict_\idxDict)\}_{\idxDict=1}^\noDict$ for each $q\in\setKernels$ is linearly independent, i.e., its corresponding kernel Gram matrix $\matKernel_q$ is of full rank, and thus $\matKernelFull$ is also linearly independent. Then, the $\matKernelFull$-norm of $\widehat{\matConsensus}$ is given by $||\widehat{\matConsensus}||_{\matKernelFull}=1$. In particular, it holds that both consensus matrices are identical to each other, i.e., $\widehat{\matConsensus}=\matConsensus$.
\end{lemma}
\begin{IEEEproof}
The proof is given in Appendix \ref{sec:proof_modConsensusNorm}.
\end{IEEEproof}
Due to Lemma \ref{lem:modConsensusMat}, for further analysis we are free to use either $\matConsensus$ or $\widehat{\matConsensus}$ and it holds that $||\matConsensus||_{\matKernelFull}=||\widehat{\matConsensus}||_{\matKernelFull}=1$.

\subsection{Convergence Analysis}
From \eqref{eq:dchypass} we can summarize both update equations of the \gls{dchypass} in terms of all coefficient vectors in the network by defining
\begin{equation}
	\vecc{z}_{\idxIter} := \left[\begin{array}{c}
					\vecWeight_{1,k} \\
					\vdots \\
					\vecWeight_{J,k}
					\end{array}\right],\,
	\vecc{y}_{\idxIter} := \left[\begin{array}{c}
						\mu_{1,\idxIter} (\vecWeight_{1,k}-P_{\hslab_{1,k}}(\vecWeight_{1,k})) \\
						\vdots \\
						\mu_{\noNodes,\idxIter} (\vecWeight_{\noNodes,k}-P_{\hslab_{\noNodes,k}}(\vecWeight_{\noNodes,k}))
						\end{array}\right]	\nonumber
\end{equation}
and rewriting \eqref{eq:localApsm} and \eqref{eq:diffStep} into 
\begin{equation}
	\vecc{z}_{\idxIter+1} = (\mat{G}\otimes\matEye_{\noDict\noKernels}) (\vecc{z}_{\idxIter}-\vecc{y}_{\idxIter}). \label{eq:dchypassFull}
\end{equation}
We show the convergence properties of \gls{dchypass} for fixed and deterministic network topologies. Although the space under study is the $\matKernel$-metric space unlike \cite{Cavalcante2009,Cavalcante2011} we can still prove the properties due to Lemmas \ref{lem:consensusMat} and \ref{lem:modConsensusMat}.

\begin{theorem} \label{theorem:analysis}
The sequence $(\vecc{z}_\idxIter)_{\idxIter\in\N}$ generated by \eqref{eq:dchypassFull} satisfies the following.
\begin{enumerate}
	\item \label{item:monotone}
	\emph{Monotone approximation:} Assume that $\vecWeightNodeTime\notin\hslabNodeTime$ with $\stepSize_{\idxNode,\idxIter}\in(0,2)$ for at least one node $\idxNode$ and that $\stepSize_{\idxNeigh,\idxIter}\in[0,2]\, (\idxNeigh\neq\idxNode)$. Then, for every $\vecWeightOpt_k\in\Upsilon_k$ and $\vecc{z}^\star_{\idxIter}:=[(\vecWeightOpt_\idxIter)^\T, (\vecWeightOpt_\idxIter)^\T, \ldots, (\vecWeightOpt_\idxIter)^\T]^\T\in\R^{\noDict\noKernels\noNodes}$ it holds that
	\begin{equation}
		||\vecc{z}_{\idxIter+1}-\vecc{z}^\star_{\idxIter}||_{\matKernelFull} < ||\vecc{z}_{\idxIter}-\vecc{z}^\star_{\idxIter}||_{\matKernelFull}
	\end{equation}
	where $\Upsilon_k\neq\emptyset$ since we assume that $\Upsilon^\star\neq\emptyset$.
\end{enumerate}
For the remaining properties we assume that $\stepSize_{\idxNode,\idxIter}\in[\epsilon_1,2-\epsilon_2]$ with $\epsilon_1,\epsilon_2>0$ and that a sufficiently large hyperslab threshold $\varepsilon_\idxNode$ per node~$\idxNode$ has been chosen such that $\vecWeightOpt\in\Upsilon^\star\neq \emptyset$. We further define $\vecc{z}^\star:=[(\vecWeightOpt)^\T, (\vecWeightOpt)^\T, \ldots, (\vecWeightOpt)^\T]^\T$. Then the following holds:
\begin{enumerate}
\setcounter{enumi}{1}
	\item \label{item:asymptotic_min}
	\emph{Asymptotic minimization of local costs:} For every $\vecc{z}^\star$ the local costs $\funCostNodeTime(\vecWeightNodeTime)=||\vecWeightNodeTime-P_{\hslabNodeTime}(\vecWeightNodeTime)||_{\matKernel}$ are asymptotically minimized, i.e.,
	\begin{equation}
	\lim_{\idxIter\to\infty} \funCostNodeTime(\vecWeightNodeTime)=0, \forall \idxNode\in\setNodes.
	\end{equation}
	
	\item \label{item:asymptotic_con}
	\emph{Asymptotic consensus:} 
	With the decomposition $\matConsensus=\mat{B}\mat{B}^\T+\mat{X}$ and $||\mat{X}||_2<1$ the sequence $(\vecc{z}_{\idxIter})_{\idxIter\in\N}$ asymptotically achieves consensus such that
	\begin{equation}
		\lim_{\idxIter\to\infty} (\matEye_{\noDict\noKernels\noNodes}-\mat{B}\mat{B}^\T)\vecc{z}_{\idxIter} = \vecc{0}.
	\end{equation}
	
	\item \label{item:convergence} 
	\emph{Convergence of $(\vecc{z}_\idxIter)_{\idxIter\in\N}$:} 
	Suppose that $\Upsilon^\star$ has a nonempty interior, i.e., there exists $\rho>0$ and interior point $\tilde{\vecc{u}}$ such that $\{ \vecc{v}\in\R^{\noDict\noKernels} \,|\, ||\vecc{v}-\tilde{\vecc{u}}||_{\matKernel} \leq \rho \}\subset\setSolutionOpt$. Then, the sequence $(\vecc{z}_\idxIter)_{\idxIter\in\N}$ converges to a vector $\widehat{\vecc{z}}=[  \widehat{\vecWeight}^\T, \ldots, \widehat{\vecWeight}^\T]^\T\in\mathcal{C}$ satisfying $(\matEye_{\noDict\noKernels\noNodes}-\mat{B}\mat{B}^\T)\widehat{\vecc{z}}=\vecc{0}$.
	
	\item \label{item:characterization}
	\emph{Characterization of limit point $\widehat{\vecc{z}}$:}
	Suppose for an interior $\tilde{\vecc{u}}\in\Upsilon^\star$ that for any $\epsilon>0$ and any $\eta>0$ there exists a $\zeta>0$ such that
	\begin{equation}
	\underset{\idxIter\in\mathcal{I}}{\min} \sum_{\idxNode\in\setNodes} ||\vecWeightNodeTime-P_{\hslabNodeTime}(\vecWeightNodeTime)||_{\matKernel} \geq \zeta,
	\end{equation}
	where
	\begin{align*}
	& \mathcal{I}:= \Big\{ \idxIter\in\N\, |\, \sum_{\idxNode\in\setNodes} \dis_{\matKernel}(\vecWeightNodeTime,\lev_{\leq 0} \funCostNodeTime) > \epsilon\\
	& \text{and } \sum_{\idxNode\in\setNodes} ||\tilde{\vecc{u}}-\vecWeightNodeTime||_{\matKernel} \leq\eta \Big\}.
	\end{align*}
	Then it holds that $\widehat{\vecWeight}\in\Upsilon$ with $\Upsilon$ defined as in \eqref{eq:bestSet}.
	
\end{enumerate}
\end{theorem}
\begin{IEEEproof}
\textcolor{black}{
The proofs of Theorem \ref{theorem:analysis}.\ref{item:monotone}-\ref{theorem:analysis}.\ref{item:asymptotic_con} can be directly deduced from the corresponding proofs of Theorem 1a)-1c) in \cite[Appendix III]{Cavalcante2009} under the consideration that $||\matConsensus||_2=||\matConsensus||_{\matKernelFull}=1$ (see Lemma \ref{lem:modConsensusMat}) and that $\funCostNodeTime(\vecWeightNode)$ is a non-negative convex function. Note that the proof of Theorem~\ref{theorem:analysis}.\ref{item:monotone} needs to be derived considering the $\matKernelFull$-metric and not the ordinary Euclidean metric as in \cite{Cavalcante2009}.
The proofs of Theorem~\ref{theorem:analysis}.\ref{item:convergence} and \ref{theorem:analysis}.\ref{item:characterization} are given in Appendix \ref{sec:proof_analysis}.
}
\end{IEEEproof}

\section{Numerical Evaluation} \label{sec:num_examples}
In the following section, we evaluate the performance of the \gls{dchypass} by applying it to the spatial reconstruction of multiple Gaussian functions, real altitude data and the tracking of a time-varying nonlinear function by a network of nodes. The nodes are distributed over the unit-square area $A$ and each node $\idxNode$ uses its Cartesian position vector $\vecInputNode=[x_{\idxNode,1}, x_{\idxNode,2}]^\T\in\spaceInput$ as its regressor. We assume that the positions of the nodes stay fixed, i.e., $\vecInputNodeTime$ does not change over time. This is not necessary for the \gls{dchypass} to be applicable, e.g. it can be applied to a mobile network where the positions change over time as investigated in \cite{Shin2018}. Per time index $\idxIter$ the nodes take a new measurement $\scaOutputNodeTime$ of the function $\funTrue$ at their position $\vecInputNode$. Hence, the network constantly monitors the function $\funTrue$. For all experiments we assume model \eqref{eq:systemModel} with zero-mean white Gaussian noise of variance $\noiseVar$. Since in this scenario measurements of the function $\funTrue$ are spatially spread over the nodes a collaboration among the nodes is inevitable for a good regression performance. Thus, it is an appropriate application example where the benefit of distributed learning becomes clear.
 
We compare the performance of the \gls{dchypass} to the \gls{rffdklms} \cite{Bouboulis2018}, the \gls{fatc-klms} \cite{Gao2015} and the \gls{mkdice} \cite{Shin2017a} which are state of the art algorithms for distributed kernel-based estimation. Both \gls{rffdklms} and \gls{fatc-klms} are single kernel approaches based on a diffusion mechanism. Assuming that the FATC-KLMS only considers local data in its adaptation step, both schemes exhibit the same number of transmissions per node as the D-CHYPASS. To enable a fair comparison we restrict the adaptation step of the \gls{fatc-klms} to use local data only and extend the algorithm by multiple kernels as in D-CHYPASS. We call this scheme the \gls{dmklms}. Its update equation per node $\idxNode$ is given by
\begin{subequations}
\begin{align}
	\vecWeight'_{j,k+1} &:= \vecWeightNodeTime+\stepSize_{\idxNode,\idxIter} \rund{ \scaOutputNodeTime-\vecWeightNodeTime^\T\vecKernel(\vecInputNodeTime) } \vecKernel(\vecInputNodeTime)\\
	\vecWeight_{j,k+1}  &:= \sum_{\idxNeigh\in\setNeighNode} g_{\idxNodeNeigh} \vecWeight'_{\idxNeigh,k+1}.
\end{align}
\end{subequations}
The \gls{rffdklms} approximates kernel evaluations by random Fourier features such that no design of a specific dictionary set is necessary. However, its performance is highly dependent on the number of the utilized Fourier features which determines the dimension of the vectors to be exchanged.
The \gls{mkdice} is a distributed regression scheme based on kernel least squares with multiple kernels using the \gls{admm} for its distributed mechanism. The number of transmissions per iteration is higher compared to the \gls{dchypass}, \gls{rffdklms} and \gls{dmklms}. Naturally, it is not an adaptive scheme but is included here for reference purposes.
As benchmark performance, we consider the central \gls{chypass} given by
\begin{equation}
	\vecWeight_{\idxIter+1} := \vecWeight_{\idxIter}-\stepSize \sum_{\idxNode\in\setNodes} \rund{ \vecWeight_{\idxIter}-P^{\matKernel}_{\hslabNodeTime}(\vecWeight_{\idxIter}) }.
\end{equation}
The central \gls{chypass} requires all node positions and measurements $\{(\vecInputNodeTime,\scaOutputNodeTime)\}_{\idxNode\in\setNodes}$ per time index $\idxIter$ at a single node to perform the projection $P^{\matKernel}_{\hslabNodeTime}(\vecWeight_{\idxIter})$ onto each set $\hslab_{\idxNode,\idxIter}$. 

Regarding the dictionaries we assume that each $\setDict_q$ uses the same samples $\{\vecDictIdx\}_{\idxDict=1}^r$. These samples are a subset of the node positions $\{\vecInputNode\}_{\idxNode\in\setNodes}$ in the network and are selected following the coherence criterion: A node position $\vecInputNode$ is compared to every dictionary entry $\{\vecDictIdx\}_{\idxDict=1}^{\noDict}$ and is included as dictionary sample $\vecDict_{\noDict+1}$ if it satisfies 
\begin{equation}
	\max_{q\in\setKernels} \max_{\idxDict={1,\ldots,\noDict}} |\kernel_q(\vecInputNode,\vecDict_\idxDict)| \leq \tau.	
\end{equation}
Here, $0<\tau\leq1$ is the coherence threshold controlling the cardinality of $\setDict_q$. The dictionary $\setDict_q$ is generated a priori over all node positions before the algorithm iterates. After that it stays fixed throughout the reconstruction process for the specific algorithm.

As error metric we consider the network $\mathrm{NMSE}_{\idxIter}$ per time $\idxIter$ over the area $A$. It evaluates the normalized squared-difference between reconstructed field $\funEstNode(\vecInput)$ and the true field $\funTrue(\vecInput)$ averaged over all nodes:
\begin{equation}
	\mathrm{NMSE}_{\idxIter} := \frac{1}{\noNodes} \sum_{\idxNode\in\setNodes} \frac{\expc{\int_A|\funTrue(\vecInput)-\vecWeightNodeTime^\T\vecKernel(\vecInput)|^2d\vecInput}}{\int_A |\funTrue(\vecInput)|^2d\vecInput}.
\end{equation}
The expectation in the numerator is approximated by averaging over independent trials. The integrals are approximated by a sum over regularly positioned grid points which sample the area $A$.

\subsection{Multiple Gaussian Functions} \label{sec:results:multi_gauss}
As a first example we apply the \gls{dchypass} algorithm to the reconstruction of two Gaussian functions with different bandwidths given as follows:
\begin{equation*}
	\funTrue(\vecInput) := 2\exp\rund{-\frac{||\vecInput-\vecc{p}_1||_{\R^2}^2}{2\cdot 0.1^2}} + \exp\rund{-\frac{||\vecInput-\vecc{p}_2||_{\R^2}^2}{2\cdot 0.3^2}}
\end{equation*}
with $\vecc{p}_1=[0.5,0.7]^\T, \vecc{p}_2=[0.3,0.1]^\T$, and the Cartesian coordinate vector  $\vecInput=[x_1,x_2]^\T$. We use $\noNodes=60$ nodes randomly placed over $A=[0,1]^2$ following a uniform distribution where nodes share a connection if their distance to each other satisfies $D<0.3$. We assume a noise variance of $\sigma_n^2=0.3$ at the nodes and average the performance over $200$ trials with a new network realization in each trial. Regarding the kernel choice we use two Gaussian kernels ($\noKernels=2$) with bandwidths $\kernelBW_1=0.1$ and $\kernelBW_2=0.3$. For all diffusion-based algorithms we use the Metropolis-Hastings weights \cite{Xiao2007} where each entry $g_{\idxNodeNeigh}$ is determined by
\begin{equation*}
	g_{\idxNodeNeigh} = 
	\begin{cases}
		\dfrac{1}{\max\{\delta_j,\delta_i\}} & \text{if }j\neq i \text{ and } (j,i)\in\setEdges \\
		1-\sum\limits_{\idxNeigh\in\setNeighNode\setminus\{\idxNode\}} \dfrac{1}{\max\{\delta_j,\delta_i\}} & \text{if } \idxNode=\idxNeigh\\
		0 & \mathrm{otherwise}
	\end{cases}
\end{equation*}
and $\delta_j=|\setNeighNode|$ denotes the degree of a node~$\idxNode$.
For all algorithms we set the coherence threshold $\tau$ such that the same average dictionary size of $\bar{\noDict}=33$ is utilized. Single kernel approaches use the arithmetic average of the bandwidths chosen for the multikernel schemes as their kernel bandwidth.
We evaluate the D-CHYPASS~(I) with a hyperplane projection, i.e., $\varepsilon_{\idxNode}=0$, and the D-CHYPASS~(II) with a hyperslab projection with $\varepsilon_{\idxNode}=0.5$.
The chosen parameter values for the considered algorithms are listed in Table \ref{tab:multi_gauss}. 

\begin{table}[tb]
\centering
\caption{Parameter values for experiment in Section \ref{sec:results:multi_gauss}}
\def\arraystretch{1.3}
\begin{tabular}{|l|l|l|l|}
\hline \textbf{Algorithm} & \multicolumn{3}{c|}{\textbf{Parameters}} \\
\hline D-CHYPASS (I) & $\stepSize_{\idxNode,\idxIter}=0.2$ & \multirow{8}{*}{$\tau=0.95$} & \\
& $\varepsilon_{\idxNode}=0$ & & \\
\cline{1-2} D-CHYPASS (II) & $\stepSize_{\idxNode,\idxIter}=0.5$ & & \\
& $\varepsilon_{\idxNode}=0.5$ & & $\kernelBW_1=0.1$\\
\cline{1-2} DMKLMS & $\stepSize_{\idxNode,\idxIter}=0.1$ & & $\kernelBW_2=0.3$\\ 
\cline{1-2} MKDiCE & $\stepSize_{\idxNode,\idxIter}=0.5$ & & \\
\cline{1-2} Central CHYPASS & $\stepSize=3.3\cdot10^{-3}$ & &\\
& $\varepsilon_{\idxNode}=0$ & & \\
\hline FATC-KLMS & $\stepSize_{\idxNode,\idxIter}=0.07$ & $\tau=0.9$ & \multirow{3}{*}{$\kernelBW=0.2$}\\
\cline{1-3} RFF-DKLMS (I) & $\stepSize_{\idxNode,\idxIter}=0.1$ & $\noDict_{\mathrm{RFF}}=100$ & \\
\cline{1-3} RFF-DKLMS (II) & $\stepSize_{\idxNode,\idxIter}=0.1$ & $\noDict_{\mathrm{RFF}}=500$ & \\
\hline 
\end{tabular}
\label{tab:multi_gauss}
\end{table}

\begin{figure}[tb]
  	\centering
  	\setlength\figureheight{4cm} 
  	\setlength\figurewidth{7cm}
	\includegraphics{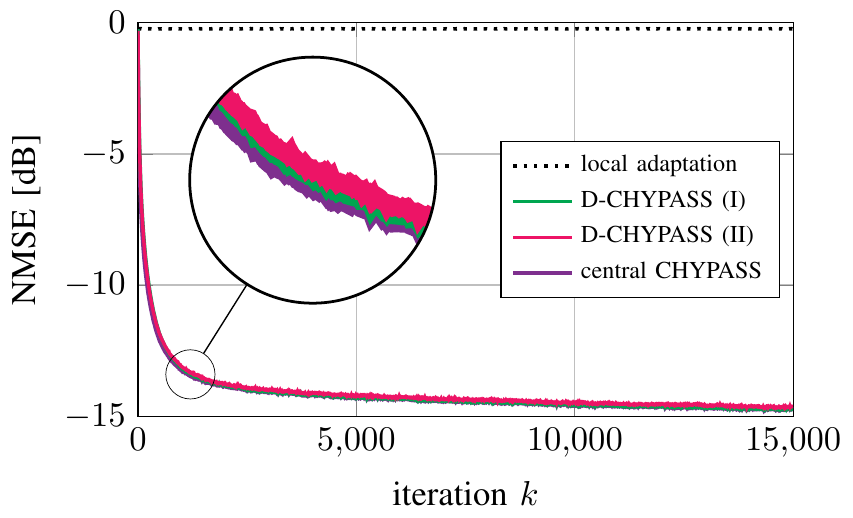}
	\caption{Comparing learning curves of \gls{dchypass} to central and local adaptation.}
	\label{fig:multi_gauss_chypass}
\end{figure}

\begin{figure}[tb]
  	\centering
  	\setlength\figureheight{4cm} 
  	\setlength\figurewidth{7cm}
	\includegraphics{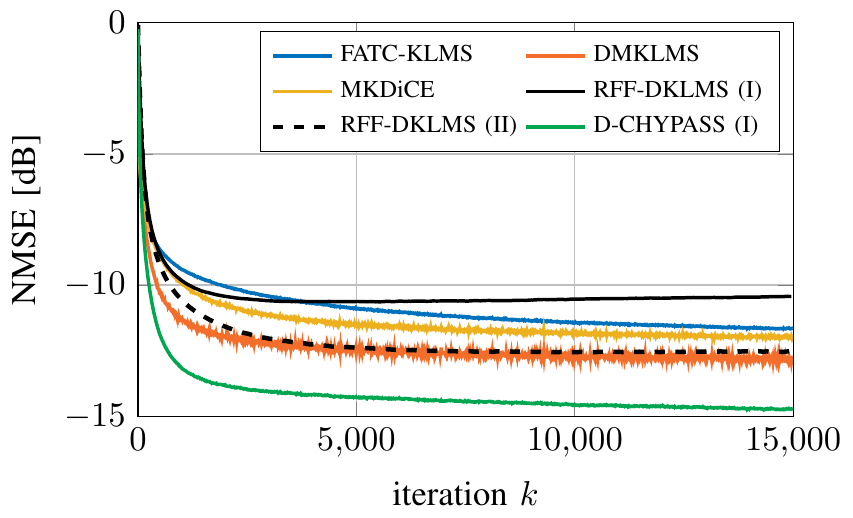}
	\caption{Learning curves for the reconstruction of multiple Gaussian functions.}
	\label{fig:multi_gauss_nmse}
\end{figure}

%

Figure~\ref{fig:multi_gauss_chypass} compares the NMSE learning curves of D-CHYPASS~(I) and D-CHYPASS~(II) to a local adaptation and the central \gls{chypass}. Clearly, the local adaptation completely fails to approximate $\funTrue$ while both D-CHYPASS~(I) and D-CHYPASS~(II) perform close to the central \gls{chypass}.
Figure~\ref{fig:multi_gauss_nmse} compares the performance of D-CHYPASS~(I) to state of the art schemes. D-CHYPASS (I) significantly outperforms the compared algorithms in terms of convergence speed and steady-state error. Regarding monokernel approaches, FATC-KLMS outperforms \gls{rffdklms}~(I) in its steady-state error although it uses a dictionary of only $\bar{\noDict}=33$ samples compared to $\noDict_{\mathrm{RFF}}=100$ random Fourier features. By increasing the number of Fourier features to $\noDict_{\mathrm{RFF}}=500$ the performance can be significantly improved, cf. \gls{rffdklms}~(II). Nevertheless, this improvement comes with a huge increase in communication overhead since the number of Fourier features is equal to the dimension of the coefficient vectors to be exchanged. While \gls{dmklms} exchanges vectors with $\bar{\noDict}\noKernels=66$ entries only, the coefficient vectors in \gls{rffdklms}~(II) have $\noDict_{\mathrm{RFF}}=500$ entries. Thus, by relying on an a priori designed dictionary as in \gls{dmklms} and \gls{dchypass}, huge savings in communication overhead and computational complexity can be achieved.
The enhanced performance by D-CHYPASS compared to the other multikernel approaches is due to a better metric in form of the $\matKernel$-norm and the normalization factor $||\matKernel^{-1}\vecKernel(\vecInputNodeTime)||_{\matKernel}^2$ in the projection $P_{\hslabNodeTime}(\vecWeightNodeTime)$ which adapts the step size $\stepSize_{\idxNode,\idxIter}$. 
By exploiting the projection \gls{wrt} the $\matKernel$-norm the shape of the cost function $\funCostNodeTime(\vecWeightNodeTime)$ is changed such that convergence speed is improved \cite{Yukawa2007}.

\begin{figure}
  	\centering
  	\setlength\figureheight{4.2cm} 
  	\setlength\figurewidth{6.5cm}
	\includegraphics{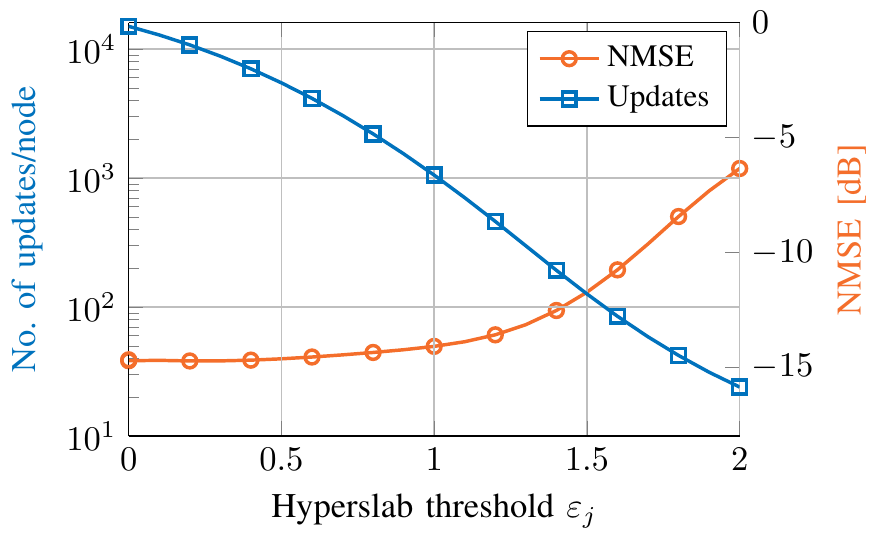}	
	\caption{\textcolor{black}{Number of updates per node and NMSE for different values of the hyperslab threshold $\varepsilon_j$ for the \gls{dchypass}.}}
	\label{fig:multi_nmseHslab}
\end{figure}

From Figure~\ref{fig:multi_gauss_chypass}, D-CHYPASS (I) and (II) show a similar performance with a negligible loss for D-CHYPASS~(II). However, this minor loss comes with a huge reduction in complexity per node~$\idxNode$. Since D-CHYPASS~(II) projects onto a hyperslab with $\varepsilon_{\idxNode}=0.5$ there is a higher probability that $\vecWeightNodeTime$ is contained in $\hslabNodeTime$ than in D-CHYPASS~(I) where $\varepsilon_{\idxNode}=0$.
If $\vecWeightNodeTime\in\hslabNodeTime$, the vector $\vecWeightNodeTime$ is not updated saving a significant amount of computations. In contrast, when using a hyperplane $(\varepsilon_{\idxNode}=0)$ the vector $\vecWeightNodeTime$ has to be updated in each iteration.
Figure~\ref{fig:multi_nmseHslab} shows the number of local APSM updates \eqref{eq:localApsm} per node in logarithmic scale over the hyperslab threshold $\varepsilon_{\idxNode}$. Additionally, the NMSE averaged over the last 200 iterations in relation to the threshold is depicted. For thresholds $\varepsilon_{\idxNode}>0$ a step size of $\stepSize_{\idxNode,\idxIter}=0.5$ is used. We can observe that using hyperslab thresholds up to $\varepsilon_{\idxNode}=0.5$ saves a huge amount of complexity while keeping the error performance constant. E.g. for D-CHYPASS~(II) with $\varepsilon_{\idxNode}=0.5$ in average 5,468 updates are executed per node. Compared to 15,000 updates for D-CHYPASS~(I) a reduction of approximately $64\%$ in computations can be achieved. This is crucial especially for sensors with low computational capability and limited battery life. However, from Figure~\ref{fig:multi_nmseHslab} it is also clear that the computational load cannot be arbitrarily reduced without degrading the reconstruction performance. This is visible especially for thresholds $\varepsilon_{\idxNode}>1$.

\begin{figure}
  	\centering
  	\setlength\figureheight{3cm} 
  	\setlength\figurewidth{7cm}
	\includegraphics{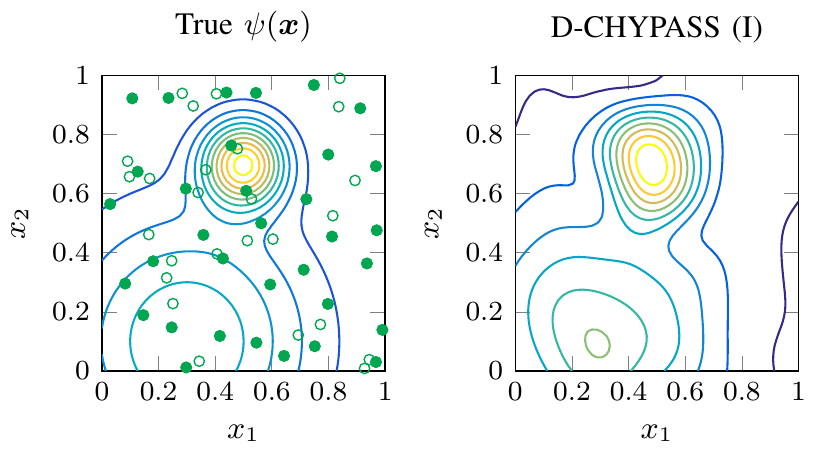}
	\caption{Contour plots of the true $\funTrue(\vecInput)$ (left) and its reconstruction $\funEst(\vecInput)$ (right) at one node using the \gls{dchypass} at steady state. Green circles show the node positions and filled circles the chosen dictionary entries.}
	\label{fig:multi_gauss_contour}
\end{figure}

\begin{figure}
  	\centering
  	\setlength\figureheight{4.4cm} 
  	\setlength\figurewidth{7cm}
	\includegraphics{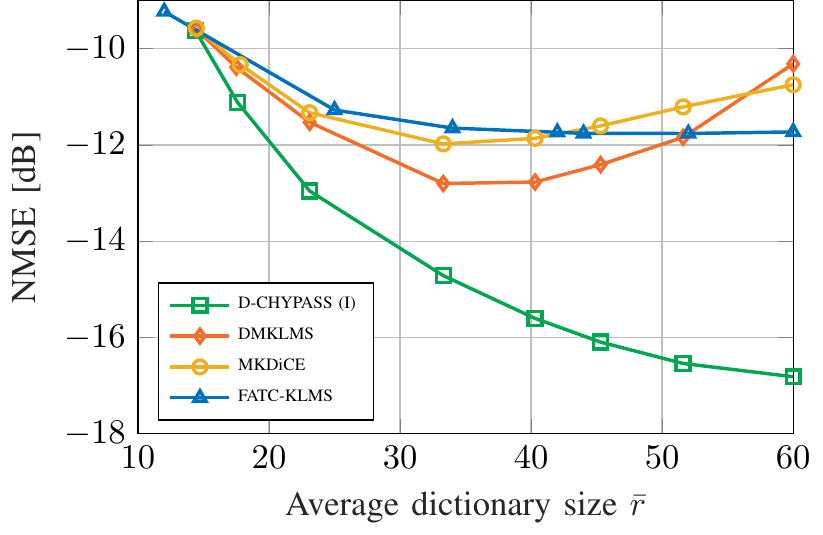}
	\caption{NMSE over dictionary size for the reconstruction of multiple Gaussian functions.}
	\label{fig:multi_gauss_nmseDictSize}
\end{figure}

\begin{figure}
  	\centering
  	\setlength\figureheight{4cm} 
  	\setlength\figurewidth{7cm}
	\includegraphics{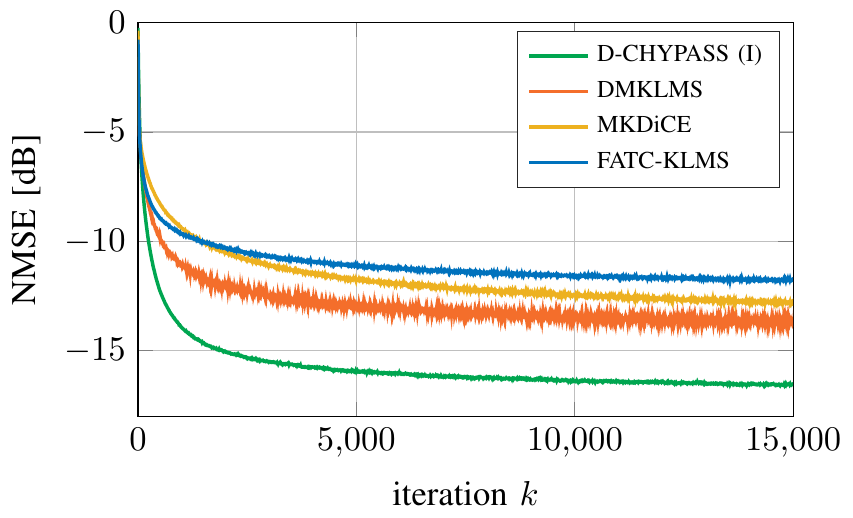}
	\caption{Learning curves for the reconstruction of multiple Gaussian functions for a coherence threshold $\tau=0.99$ corresponding to an average dictionary size of $\bar{\noDict}=53$ samples.}
	\label{fig:multi_gauss_nmse_tau99}
\end{figure}

In Figure \ref{fig:multi_gauss_contour} we depict the contour plot of the true function $\funTrue(\vecInput)$ together with an exemplary set of node positions and the reconstructed function $\funEst(\vecInput)$ by \gls{dchypass}~(I). The reconstruction is shown for one node in the network at steady state. By virtue of the consensus averaging step each node in the network will have the same reconstruction. We can observe that both Gaussian functions are approximated with good accuracy. The peaks of both functions can be clearly distinguished. However, at outer regions some inaccuracies can still be seen. These are expected to be reduced when increasing the dictionary size.

Figure \ref{fig:multi_gauss_nmseDictSize} shows the error performance of the algorithms over the averaged dictionary size $\bar{\noDict}$ for 200 trials. The NMSE values are calculated as an average over the last 200 iterations with again 15,000 iterations for each algorithm. We observe that \gls{dchypass} (I) outperforms its competitors with growing dictionary size. In particular, \gls{dmklms} and \gls{mkdice} lose in performance for dictionary sizes $\bar{\noDict}>33$ while \gls{dchypass} steadily improves its reconstruction. This is due to the reason that for \gls{dmklms} and \gls{mkdice} the step size has to be adjusted to the growing dictionary size to avoid an increasing steady-state error. In \gls{dmklms} the step size is not normalized to the squared norm of the kernel vector $\vecKernel(\vecInput)$ as in \gls{dchypass} which can lead to divergence. 
Regarding FATC-KLMS a similar effect is expected to appear for higher dictionary sizes $\bar{\noDict}>60$ since it uses one kernel only. Therefore, in the range of 40 to 60 dictionary samples it performs better than \gls{dmklms} and \gls{mkdice}.
To show that the performance of \gls{mkdice} and \gls{dmklms} can be improved, in Figure \ref{fig:multi_gauss_nmse_tau99} we depict the NMSE performance for an adapted step size over the iteration with $\tau=0.99$. This coherence threshold results in an average dictionary size of $\bar{\noDict}=53$, a point where \gls{mkdice} and \gls{dmklms} show degrading performance according to Figure~\ref{fig:multi_gauss_nmseDictSize}. The step sizes are chosen as $\stepSize_{\idxNode,\idxIter}=0.05$ for \gls{dmklms} and $\stepSize_{\idxNode,\idxIter}=0.2$ for \gls{mkdice}, respectively. We observe that by adjusting the step size to the dictionary size the steady-state performance at $\bar{\noDict}=53$ is improved compared to Figure \ref{fig:multi_gauss_nmseDictSize}. Now, both \gls{mkdice} and \gls{dmklms} outperform FATC-KLMS.
\begin{remark}
Regarding \gls{dchypass} (I) it should be noted that for $\tau>0.98$ divergence was observed. This is caused by the inversion of an ill-conditioned kernel Gram matrix $\matKernel$. This occurs if the dictionary employs node positions close to each other leading to linear dependency in $\matKernel$. With a higher coherence threshold the probability of such a case increases. To numerically stabilize the inversion of $\matKernel$ a scaled identity matrix $\gamma \matEye_{\noDict\noKernels}$ is added to the matrix as regularization. The matrix $\matKernel$ in \eqref{eq:projection_slab} is then substituted by $\matKernel+\gamma\matEye_{\noDict\noKernels}$. For thresholds $\tau>0.98$ a regularization parameter of $\gamma=0.01$ was used in this experiment to achieve a stable performance.
\end{remark}


\subsection{Real Altitude Data} \label{sec:results:real_data}
We apply \gls{dchypass} to the reconstruction of real altitude data where each node measures the altitude at its position $\vecInputNode$. For the data we use the ETOPO1 global relief model which is provided by the National Oceanic and Atmospheric Administration \cite{Amante2009} and which exhibits several low/high frequency components. In the original data the position is given by the longitude and latitude and the corresponding altitude $\funTrue(\vecInput)$ is delivered for each such position. As in \cite{Toda2017}, we choose an area of $31\times 31$ points with longitudes $\{138.5,138.5+\frac{1}{60},\ldots,139\}$ and latitudes $\{34.5,34.5+\frac{1}{60},\ldots,35\}$. However, for easier handling we map longitudes and latitudes to Cartesian coordinates in the unit-square area such that $\vecInput\in[0,1]^2$. We consider $\noNodes=200$ randomly placed over the described area. Nodes with a distance $D<0.2$ to each other share a connection. We assume noise with $\noiseVar=0.3$. 
The coherence threshold is set such that each algorithm employs a dictionary of average size $\bar{\noDict}=105$ while the RFF-DKLMS uses $\noDict_{\mathrm{RFF}}=200$ Fourier features.
The performances are averaged over 200 independent trials. Table \ref{tab:izu} lists the chosen parameter values for the considered algorithms.

\begin{table}[tb]
\centering
\caption{Parameter values for experiment in section \ref{sec:results:real_data}}
\def\arraystretch{1.3}
\begin{tabular}{|l|l|l|l|}
\hline \textbf{Algorithm} &  \multicolumn{3}{c|}{\textbf{Parameters}} \\
\hline D-CHYPASS & $\stepSize_{\idxNode,\idxIter}=0.5,\varepsilon_{\idxNode}=0$ & \multirow{3}{*}{$\tau=0.85$} &  $\kernelBW_1=0.06$\\ 
\cline{1-2} DMKLMS & $\stepSize_{\idxNode,\idxIter}=0.05$ & &$\kernelBW_2=0.1$ \\
\cline{1-2} MKDiCE & $\stepSize_{\idxNode,\idxIter}=0.7$ & &  \\
\hline FATC-KLMS & $\stepSize_{\idxNode,\idxIter}=0.05$ & $\tau=0.78$ & $\kernelBW=0.08$\\
\hline RFF-DKLMS & $\stepSize_{\idxNode,\idxIter}=0.2$ & $\noDict_{\mathrm{RFF}}=200$ & $\kernelBW=0.08$\\
\hline 
\end{tabular}
\label{tab:izu}
\end{table}

\begin{figure}[tb]
  	\centering
  	\setlength\figureheight{4cm} 
  	\setlength\figurewidth{7cm}
	\includegraphics{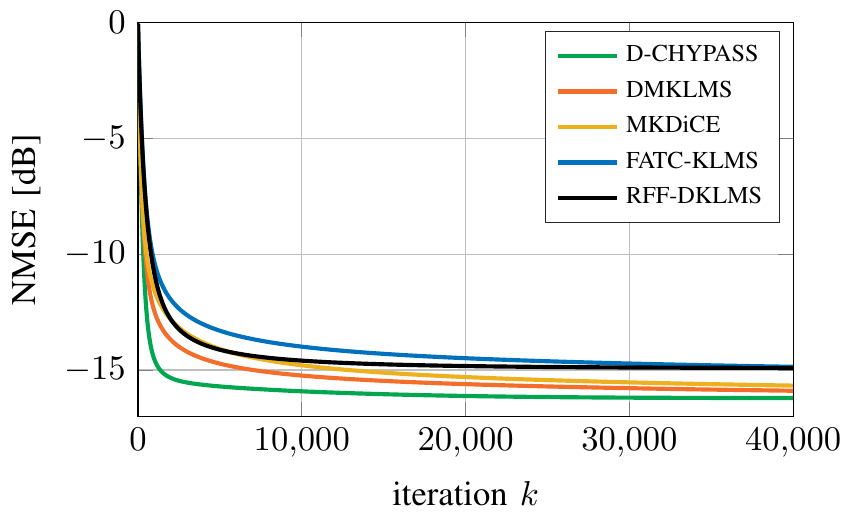}
	\caption{Learning curves for the reconstruction of altitude data.}
	\label{fig:izu_nmse}
\end{figure}

\begin{figure}[tb]
  	\centering
  	\setlength\figureheight{5cm} 
  	\setlength\figurewidth{7cm}
	\includegraphics{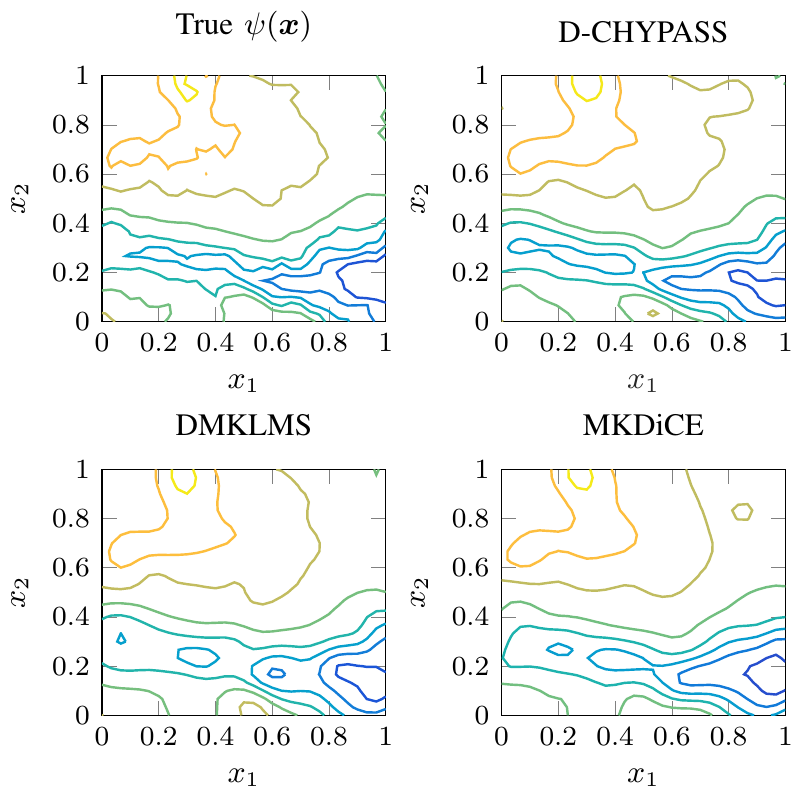}
	\caption{Contour plots of the altitude reconstruction by one node for the D-CHYPASS, DMKLMS, and MKDiCE.}
	\label{fig:izu_contour}
\end{figure}

Figure \ref{fig:izu_nmse} depicts the NMSE performance over the iteration. Again \gls{dchypass} outperforms the other algorithms in terms of convergence speed and steady-state error. Although \gls{dmklms} performs very close to D-CHYPASS it can be observed that the convergence speed of D-CHYPASS is faster. FATC-KLMS and \gls{rffdklms} perform worst since their reconstruction capability is limited by the use of one kernel only. While \gls{rffdklms} converges faster than FATC-KLMS it should be noted that it produces a higher communication overhead due to the use of $\noDict_{\mathrm{RFF}}=200$ Fourier features compared to $\bar{\noDict}=105$ dictionary samples in FATC-KLMS.
The contour plots for the multikernel approaches at steady-state at one node are shown in Figure \ref{fig:izu_contour}. For the \gls{dchypass} we can observe a good reconstruction of the original $\funTrue(\vecInput)$ although details in the area around $[0.4, 0.7]^\T$ and $[0.4, 0.3]^\T$ are missing. The reconstructions by \gls{dmklms} and \gls{mkdice} show a less accurate approximation especially in the areas around the valley $[0.4,0.3]^\T$.

\subsection{Time-Varying Nonlinear Function} \label{sec:results:dynamic}
In the following, we examine the tracking performance of the \gls{dchypass} \gls{wrt} time-varying functions. To this end, we consider the following function being dependent on both the position $\vecc{x}$ and time $\idxIter$:
\begin{align*}
	\funTrue(\vecInput,\idxIter) &= 0.8\exp\rund{-\frac{||\vecInput-\vecc{p}_1||_{\R^2}^2}{2(1-0.5\sin(2\pi 10^{-3} \idxIter))\cdot 0.3^2}} \\
	&+\exp\rund{-\frac{||\vecInput-\vecc{p}_2||_{\R^2}^2}{2(1+0.5\sin(2\pi 10^{-3} \idxIter))\cdot 0.1^2}}
\end{align*}
with $\vecc{p}_1=[0.6, 0.5]^\T$ and $\vecc{p}_2=[0.25, 0.3]^\T$. This function contains two Gaussian shapes whose bandwidths are expanding and shrinking over time $\idxIter$. We apply the \gls{dchypass} to the reconstruction of the time-varying function $\funTrue(\vecInput,\idxIter)$ and compare it to the \gls{mkdice} and \gls{dmklms}. We use a network of $\noNodes=80$ nodes randomly distributed over the unit-square area and average the performance over 200 trials with a new network realization in each trial. The noise variance is $\noiseVar=0.3$. 
For the considered algorithms we set $\tau$ such that an average dictionary size of $\bar{\noDict}=36$ samples is achieved. We evaluate the D-CHYPASS with one and two kernels.
Table \ref{tab:dynamic} lists the chosen parameter values for the considered algorithms.

\begin{figure}[tb]
  	\centering
  	\setlength\figureheight{4cm} 
  	\setlength\figurewidth{7cm}
	\includegraphics{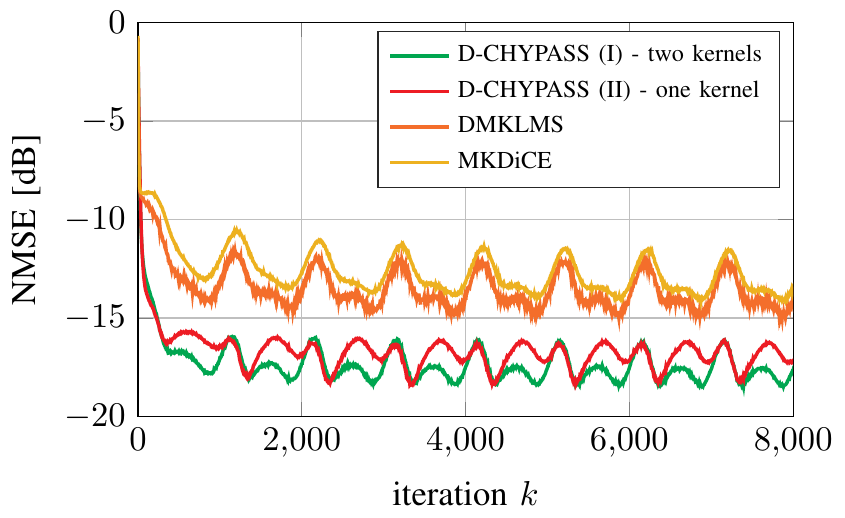}
	\caption{NMSE performance over iteration number for the tracking of a time-varying function.}
	\label{fig:dynamic_nmse}
\end{figure}

Figure \ref{fig:dynamic_nmse} shows the NMSE over the iteration number $\idxIter$. The fluctuations in the error curves are due to the time-varying bandwidths in $\funTrue(\vecInput,\idxIter)$. For all algorithms these fluctuations stay in a specific error range illustrating that the function $\funTrue(\vecInput,\idxIter)$ can be tracked within a certain range of accuracy. We observe that D-CHYPASS (I) and (II) significantly outperform the remaining algorithms. Additionally, the range of the fluctuations in the error is lower for D-CHYPASS compared to the other algorithms. It is also visible that utilizing two kernels in D-CHYPASS~(I) improves the tracking performance compared to using one kernel as in D-CHYPASS~(II). Nevertheless it is worth noting, that even with only one kernel the D-CHYPASS~(II) outperforms the multikernel approaches DMKLMS and MKDiCE illustrating the significant gain by employing the $\matKernel$-norm in the algorithm.

\begin{table}[tb]
\centering
\caption{Parameter values for experiment in section \ref{sec:results:dynamic}}
\def\arraystretch{1.3}
\begin{tabular}{|l|lll|}
\hline \textbf{Algorithm} & \multicolumn{3}{c|}{\textbf{Parameters}} \\
\hline D-CHYPASS (I) & $\stepSize_{\idxNode,\idxIter}=0.5$ & $\tau=0.95$ & $\kernelBW_1=0.1,\kernelBW_2=0.3$ \\
& $\varepsilon_{\idxNode}=0$ & &\\
\hline D-CHYPASS (II) & $\stepSize_{\idxNode,\idxIter}=0.5$ & $\tau=0.9$ & $\kernelBW_1=0.2$ \\
& $\varepsilon_{\idxNode}=0$ & & \\
\hline DMKLMS & $\stepSize_{\idxNode,\idxIter}=0.1$ & $\tau=0.95$ & $\kernelBW_1=0.1,\kernelBW_2=0.3$ \\
\hline MKDICE & $\stepSize_{\idxNode,\idxIter}=0.5$ & $\tau=0.95$ & $\kernelBW_1=0.1,\kernelBW_2=0.3$ \\
\hline 
\end{tabular}
\label{tab:dynamic}
\end{table}

\subsection{Computational Complexity and Communication Overhead}
\begin{figure}[htb]
  	\centering
  	\setlength\figureheight{3cm} 
  	\setlength\figurewidth{7.2cm}
	\includegraphics{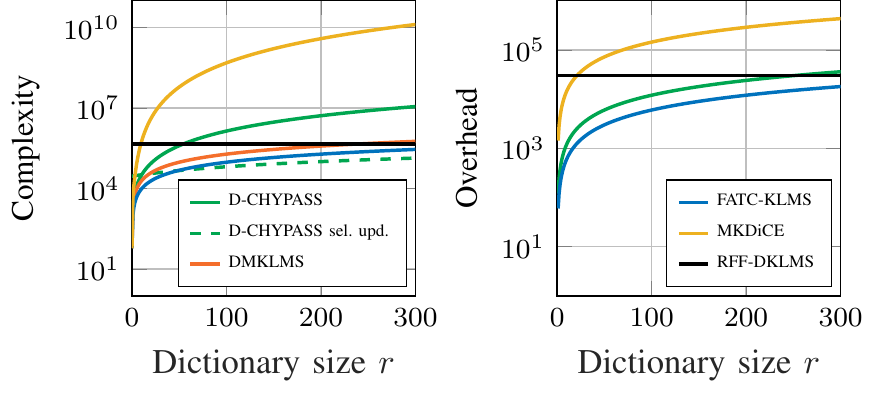}
	\caption{Computational complexity and communication overhead of the algorithms per iteration $\idxIter$ over the dictionary size~$\noDict$.}
	\label{fig:complexity}
\end{figure}

We analyze the complexities and communication overhead of the algorithms per iteration in the network. For the complexities we consider the number of multiplications and assume that Gaussian kernels are used as in \eqref{eq:gausskernel}. Furthermore, each dictionary $\setDict_q$ is designed a priori, stays fixed over time~$\idxIter$ and is common to all nodes. Therefore, the kernel Gram matrix $\matKernel$ can be computed offline before the iterative process of D-CHYPASS avoiding an inversion in each iteration. Note that $\matKernel$ is block diagonal such that $\noKernels$ inversions of $\noDict\times\noDict$ matrices have to be computed. This results in a complexity of order $\mathcal{O}(\noNodes\noKernels\noDict^3)$ in the network before D-CHYPASS starts iterating. To further reduce the complexity of D-CHYPASS the selective-update strategy can be applied which selects the $s$ most coherent dictionary samples such that only $s$ entries of the coefficient vector $\vecWeightNodeTime$ are updated \cite{Takizawa2016a}. Usually, $s\leq5$ so that $s\ll\noDict$. Then per iteration $\idxIter$ the inverse of an $s\times s$ matrix has to be computed while the complexity of the multiplications is heavily reduced. 
For the overhead we count the number of transmitted scalars among all nodes. All algorithms except the MKDiCE use a consensus averaging step which produces only broadcast transmissions. Beside broadcasts the MKDiCE comprises also unicast transmissions of vectors which depend on the receiving  node and which increase the overhead significantly.
Table~\ref{tab:complexity} lists the complexities and overhead of the algorithms where the complexity for an inversion of a $p\times p$ matrix is denoted by $v_{\text{inv}}(p):=p^3$. 
Figure \ref{fig:complexity} depicts the complexity and the overhead over the dictionary size $\noDict$ for $L=2,s=7,\noKernels=2$ and a network of $\noNodes=60$ nodes with $|\setEdges|=300$ edges. The RFF-DKLMS with $\noDict_{\text{RFF}}=500$ is included as reference. It can be clearly seen that the complexity and overhead of the ADMM-based MKDiCE are highest among the algorithms due to the inversion of a $\noKernels\noDict\times\noKernels\noDict$ matrix per iteration $\idxIter$ and the transmission of unicast vectors, respectively. Furthermore, for dictionary sizes up to $\noDict=50$ the D-CHYPASS has lower complexity than the RFF-DKLMS. By including the selective-update strategy the complexity of D-CHYPASS is significantly reduced and is even lower than single kernel FATC-KLMS.
D-CHYPASS and DMKLMS exhibit the same overhead per iteration which is lower compared to that of RFF-DKLMS for dictionary sizes up to $\noDict=200$. 
\begin{table}[htb]
\centering
\caption{Computational complexity and overhead of algorithms}
\def\arraystretch{1.3}
\begin{tabular}{|c|c|c|}
\hline \textbf{Algorithm} & \textbf{Complexity} & \textbf{Overhead}\\
\hline D-CHYPASS & $\rund{2|\setEdges|+\noNodes(L+4)}\noKernels\noDict$ & \multirow{5}{*}{$\noNodes\noKernels\noDict$} \\
& $+ (\noKernels\noDict^2+2)\noNodes$ & \\
\cline{1-2} D-CHYPASS & $\rund{(L+1)\noKernels\noDict+v_{\text{inv}}(s)+s^2+2}\noNodes$ &\\
(selective update) & $+(2|\setEdges|+3\noNodes)s$ & \\
\cline{1-2}  DMKLMS & $\rund{2|\setEdges|+\noNodes(L+4)}\noKernels\noDict+\noNodes$ & \\
\hline FATC-KLMS & $\rund{2|\setEdges|+\noNodes(L+4)}\noDict+\noNodes$ & $\noNodes\noDict$\\
\hline MKDiCE & $\rund{6|\setEdges|+4\noNodes+L+2}\noKernels\noDict$ & $2\noNodes\noKernels\noDict$\\
 &$+\noNodes\rund{1+(\noKernels\noDict)^2+v_{\text{inv}}(\noKernels\noDict)}$ & $+2|\setEdges|\noKernels\noDict$ \\
\hline RFF-DKLMS & $\noNodes(4\noDict_{\text{RFF}}+1)+(2|\setEdges|+\noNodes)\noDict_{\text{RFF}}$ & $\noNodes\noDict_{\text{RFF}}$ \\
\hline
\end{tabular}
\label{tab:complexity}
\end{table}

%

\section{Conclusion}
We proposed an adaptive learning algorithm exploiting multiple kernels and projections onto hyperslabs for the regression of nonlinear functions in diffusion networks. 
We provided a thorough convergence analysis regarding monotone approximation, asymptotic minimization, consensus and the limit point of the algorithm. To this end, we introduced a novel modified consensus matrix which we proved to be identical to the ordinary consensus matrix.
As an application example we investigated the proposed scheme for the reconstruction of spatial distributions by a network of nodes with both synthetic and real data. Note that it is not restricted to such a scenario and can be applied in general to any distributed nonlinear system identification task. Compared to the state of the art algorithms we could observe significant gains in error performance, convergence speed and stability over the employed dictionary size. In particular, our proposed \gls{apsm}-based algorithm significantly outperformed an \gls{admm}-based multikernel scheme (MKDiCE) in terms of error performance with highly decreased complexity and communication overhead. By embedding the hyperslab projection the computational demand per node could be drastically reduced over a certain range of thresholds while keeping the error performance constant.

\appendices

\section{Derivation in Cartesian Product Space} \label{sec:appendix_funSpace}
We equivalently formulate problem \eqref{eq:opt_problem_parameter} in the Cartesian product space of $\noKernels$ \gls{rkhs}s. Furthermore, we derive the local APSM update \eqref{eq:localAPSM} exploiting the isomorphism between product space and Euclidean space using the $\matKernel$-metric.
\subsection{Equivalent Problem Formulation}
\textcolor{black}{
Since $\funTrue$ lies in the sum space $\spaceSum$ of $\noKernels$ \gls{rkhs}s it is decomposable into the sum $\funTrue:=\sum_{q\in\setKernels} \funTrue_{(q)}$ where $\funTrue_{(q)}\in\spaceRKHS_q$. Thus, to estimate $\funTrue$ one approach is to minimize the metric distance between the estimate $\funEst$ and the true function $\funTrue$ in the sum space $\spaceSum$. The estimate $\funEst$ is a multikernel adaptive filter \eqref{eq:funMkaf} and can be expressed as a decomposable sum
$\funEst := \sum_{q\in\setKernels} \funEst_{(q)} \in \spaceSum$
with $\funEst_{(q)} := \sum_{\idxDict=1}^\noDict \scaWeight_{q,\idxDict} \kernel_q(\cdot,\vecDictIdx)$.
The problem can therefore be formulated as the following functional optimization problem:
\begin{equation}
	\min_{\funEst\in\spaceSum} ||\funEst-\funTrue||_{\spaceSum}.
	\label{eq:genFuncProblem}
\end{equation}
However, in the sum space $\spaceSum$ the norm and inner product have no closed-form expressions and functions might not be uniquely decomposable depending on the choice of kernel functions. An alternative approach is to formulate the problem in the Cartesian product space $\spaceProd$, in which functions are uniquely decomposable independent of the underlying kernel functions and its norm has the closed-form expression \cite{Yukawa2015}
\begin{equation}
	||F||_{\spaceProd}:= \sqrt{ \sum_{q\in\setKernels} ||f_{(q)}||_{\spaceRKHS_q}^2},\, \forall F:=(f_{(q)})_{q\in\setKernels}\in\spaceProd.
\end{equation}
Instead of the sum $\funEst=\sum_{q\in\setKernels} \funEst_{(q)}$ we consider the $Q$-tuple $\Phi:=(\funEst_{(q)})_{q\in\setKernels}$ in $\spaceProd$ with $\Phi:\spaceInput\to\R^\noKernels$. In the same way for $\funTrue$ we consider the $Q$-tuple $\Psi:=(\funTrue_{(q)})_{q\in\setKernels}$.
Furthermore, each monokernel filter $\funEst_{(q)}$ is in fact an element of the \emph{dictionary subspace} $\spaceDict_q:=\mathrm{span} \{ \kernel_q(\cdot,\vecDictIdx)\}_{\idxDict=1}^\noDict$. Thus, $\Phi$ lies in the Cartesian product of $\noKernels$ dictionary subspaces
\begin{equation}
	\spaceDict^\times:=\spaceDict_1 \times \spaceDict_2 \times \ldots \times \spaceDict_\noKernels \subset\spaceProd.
\end{equation}
With these considerations we formulate the problem in $\spaceProd$:
\begin{equation}
	\min_{\Phi\in\spaceDict^\times} ||\Phi-\Psi||_{\spaceProd}.
	\label{eq:funcProblem}
\end{equation}
The solution to the above problem is directly given by the projection $P_{\spaceDict^\times}(\Psi)$ of $\Psi$ onto the dictionary subspace $\spaceDict^\times$. This projection is in fact the best approximation of $\Psi$ in $\spaceDict^\times$ in the $\spaceProd$-norm sense. For the monokernel case $\noKernels=1$ it has been shown that under certain assumptions $P_{\spaceDict^\times}(\Psi)$ approximately equals the \gls{mmse} estimate \cite{Yukawa2016}. Exact equality to the \gls{mmse} estimate can be achieved when operating the learning procedure in the space of square-integrable functions under the probability measure (usually denoted as $L^2$ space) \cite{Ohnishi}. 
}

\textcolor{black}{
With respect to the network we denote the estimate of each node~$\idxNode$ by $\Phi_{\idxNode}$.
Then the solution of \eqref{eq:funcProblem} needs to be approached by each $\Phi_{\idxNode}$. To this end, we design the hyperslab $\widetilde{\hslab}_{\idxNode,\idxIter} := \{\Phi\in\spaceDict^{\times} : |\langle\Phi,\kernel(\cdot,\vecInputNodeTime)\rangle_{\spaceRKHS^\times}-\scaOutputNodeTime| \leq \varepsilon_j\}$ for each node~$\idxNode$ containing all $\Phi$s of bounded local instantaneous error at time instant~$\idxIter$.
Here, $\kernel:\spaceInput\times\spaceInput\to\spaceOutput^\noKernels, (\vecInput,\vecInput') \mapsto (\kernel_q(\vecInput,\vecInput'))_{q\in\setKernels}$ with $\vecInput,\vecInput'\in\spaceInput$ and $\kernel(\cdot,\vecInput) := \rund{\kernel_q(\cdot,\vecInput)}_{q\in\setKernels}\in\spaceProd$.
The estimated output for $\vecInputNodeTime$ is then given by $\funEst(\vecInputNodeTime)= \langle\Phi,\kernel(\cdot,\vecInputNodeTime)\rangle_{\spaceRKHS^\times}=\sum_{q\in\setKernels} \inprod{\funEst_{(q)},\kernel_q(\cdot,\vecInputNodeTime)}_{\spaceRKHS_q}$ using the reproducing property of each kernel. It is reasonable to assume that with high probability each set $\widetilde{\hslab}_{\idxNode,\idxIter}$ contains the corresponding \gls{mmse} estimate and thus also the optimal solution of \eqref{eq:funcProblem}, cf. \cite{Yamada2002}.
Hence, considering the stochastic property sets $\widetilde{\hslab}_{\idxNode,\idxIter}$ for all $\noNodes$ nodes and arbitrarily many time instants $\idxIter\geq 0$ their intersection $\bigcap_{\idxIter\geq0} \bigcap_{\idxNode\in\setNodes} \widetilde{\hslab}_{\idxNode,\idxIter}$ will contain the \gls{mmse} estimate as well. The objective must be to approach this intersection by the estimate $\Phi_\idxNode$ of each node~$\idxNode$. The \gls{apsm} framework can be used to approach such an intersection. To this end, we define the local cost function $\widetilde{\Theta}_{\idxNode,\idxIter}$ per node~$\idxNode$ as the metric distance between the estimate $\Phi_{\idxNode}$ and its projection $P^{\spaceRKHS^{\times}}_{\widetilde{\hslab}_{\idxNode,\idxIter}}(\Phi_{\idxNode})$ onto the hyperslab $\widetilde{\hslab}_{\idxNode,\idxIter}$ in $\spaceRKHS^\times$:
\begin{equation}
	\widetilde{\Theta}_{\idxNode,\idxIter}(\Phi_{\idxNode}) = ||\Phi_{\idxNode}-P^{\spaceProd}_{\widetilde{\hslab}_{\idxNode,\idxIter}}(\Phi_{\idxNode})||_{\spaceProd}.
	\label{eq:functionalLocalCost}
\end{equation}
Then we can formulate the following global optimization problem \gls{wrt} the functions $\{\Phi_{\idxNode}\}_{\idxNode\in\setNodes}$ over the whole network:
\begin{subequations}\label{eq:opt_problem_function}
\begin{align}
	\underset{\{\Phi_{\idxNode}|\idxNode\in\setNodes\}}{\min} &\sum_{\idxNode\in\setNodes} \widetilde{\Theta}_{\idxNode,\idxIter}(\Phi_{\idxNode})\\
	\mathrm{s.t. } &\quad \Phi_{\idxNode} = \Phi_{\idxNeigh}, \quad \forall\idxNeigh\in\setNeighNode, \label{eq:consensus}
\end{align}
\end{subequations}
where \eqref{eq:consensus} enforces a consensus among all estimates $\{\Phi_{\idxNode}\}_{\idxNode\in\setNodes}$ in the network. 
Regarding the relation between product and parameter space we note that the finite-dimensional dictionary subspace $(\spaceDict^\times,\inprod{\cdot,\cdot}_{\spaceProd})$ is isomorphic to the Euclidean parameter space $(\R^{\noDict\noKernels},\inprod{\cdot,\cdot}_{\matKernel})$, see \cite[Lemma 1]{Takizawa2016a}. In the parameter space the inner product of $\spaceProd$ is preserved by $\inprod{\cdot,\cdot}_{\matKernel}$ and $\Phi_\idxNode$ is equivalent to the respective coefficient vector $\vecWeightNode\in\R^{\noDict\noKernels}$.
Then, under the correspondence $\spaceDict^\times \ni \Phi_{\idxNode} \longleftrightarrow \vecWeightNode \in\R^{\noDict\noKernels}$ 
problem \eqref{eq:opt_problem_function} is the equivalent formulation in the product space $\spaceProd$ of problem \eqref{eq:opt_problem_parameter} such that $\widetilde{\Theta}_{\idxNode,\idxIter}(\Phi_{\idxNode}) = \Theta_{\idxNode,\idxIter}(\vecWeightNode)$ holds.
}

\subsection{Derivation of Local APSM Update}
\textcolor{black}{
For the derivation of the local update in \gls{dchypass} based on \eqref{eq:opt_problem_function} we first consider the local \gls{apsm} update for $\Phi_{\idxNode,\idxIter}$ under the assumption that $\Phi_{\idxNode,\idxIter}\notin\widetilde{\hslab}_{\idxNode,\idxIter}$:
\begin{equation}
	\Phi'_{\idxNode,\idxIter+1} = \Phi_{\idxNode,\idxIter} - \stepSize_{\idxNode,\idxIter} \frac{\widetilde{\Theta}_{\idxNode,\idxIter}(\Phi_{\idxNode,\idxIter})-\widetilde{\funCost}_{\idxNode,\idxIter}^\star}{||\widetilde{\Theta}'_{\idxNode,\idxIter}(\Phi_{\idxNode,\idxIter})||_{\spaceRKHS^{\times}}^2}  \widetilde{\Theta}'_{\idxNode,k}(\Phi_{\idxNode,\idxIter})
	\label{eq:apsmFunc}
\end{equation}
A subgradient of $\widetilde{\Theta}_{\idxNode,\idxIter}(\Phi_{\idxNode,\idxIter})$ is given by
\begin{equation}
	\widetilde{\funCost}'_{\idxNode,k}(\Phi_{\idxNode,\idxIter}):= \frac{\Phi_{\idxNode,\idxIter}-P^{\spaceRKHS^{\times}}_{\widetilde{\hslab}_{\idxNode,\idxIter}}(\Phi_{\idxNode,\idxIter})}{||\Phi_{\idxNode,\idxIter}-P^{\spaceRKHS^{\times}}_{\widetilde{\hslab}_{\idxNode,\idxIter}}(\Phi_{\idxNode,\idxIter})||_{\spaceRKHS^{\times}}}.
	\label{eq:funSubgradient}
\end{equation}
Inserting \eqref{eq:functionalLocalCost} and \eqref{eq:funSubgradient} into the APSM update \eqref{eq:apsmFunc} we achieve
\begin{equation}
	\Phi'_{\idxNode,\idxIter+1} = \Phi_{\idxNode,\idxIter} - \stepSize_{\idxNode,\idxIter} \rund{ \Phi_{\idxNode,\idxIter}-P_{\widetilde{\hslab}_{\idxNode,\idxIter}}^{\spaceRKHS^{\times}}(\Phi_{\idxNode,\idxIter}) }.
\end{equation}
The projection $P_{\widetilde{\hslab}_{\idxNode,\idxIter}}^{\spaceRKHS^{\times}}(\Phi_{\idxNode,\idxIter})$ can be calculated by \cite{Yukawa2015}
\begin{align}
	P_{\widetilde{\hslab}_{\idxNode,\idxIter}}^{\spaceRKHS^{\times}}(\Phi) =
	\begin{dcases}
	\Phi, \quad\quad \text{if } \Phi\in \widetilde{\hslab}_{\idxNode,\idxIter}\\
	\Phi-\dfrac{\Phi(\vecInput_{\idxNode,\idxIter})-\scaOutputNodeTime-\varepsilon_j}{\sum_{q\in\setKernels}||P_{\spaceDict_q}(\kernel_q(\cdot,\vecInputNodeTime))||_{\spaceRKHS_q}^2}\\\times\sum_{q\in\setKernels}P_{\spaceDict_q}(\kernel_q(\cdot,\vecInputNodeTime)),\\ \quad \text{if} \quad
	 \Phi(\vecInputNodeTime)>\scaOutputNodeTime+\varepsilon_j \\ 
	\Phi-\dfrac{\Phi(\vecInputNodeTime)-\scaOutputNodeTime+\varepsilon_j}{\sum_{q\in\setKernels}||P_{\spaceDict_q}(\kernel_q(\cdot,\vecInputNodeTime))||_{\spaceRKHS_q}^2} \\\times\sum_{q\in\setKernels}P_{\spaceDict_q}(\kernel_q(\cdot,\vecInputNodeTime)),\\ \quad \text{if} \quad \Phi(\vecInputNodeTime)<\scaOutputNodeTime-\varepsilon_j
	\end{dcases}
	\label{eq:functionalProjection}
\end{align}
where $P_{\spaceDict_q}$ is the projection operator onto the dictionary subspace $\spaceDict_q$. By virtue of Lemma 1 in \cite{Yukawa2015} it holds that $P_{\spaceDict_q}(\kernel_q(\cdot,\vecInputNodeTime))=\sum_{\idxDict=1}^{\noDict} \alpha_{\idxNode,\idxDict}^{(q)} \kernel_q(\vecDictIdx,\cdot)$ with the coefficients $\vecc{\alpha}_{\idxNode}^{(q)} := [\alpha_{\idxNode,1}^{(q)},\ldots,\alpha_{\idxNode,\noDict}^{(q)}]^\T = \matKernel_q^{-1}\vecKernel_q(\vecInputNodeTime)$.
Examining the squared norm in the denominator of the projection $P_{\widetilde{\hslab}_{\idxNode,\idxIter}}^{\spaceRKHS^{\times}}(\Phi)$ yields
\begin{align*}
	||P_{\spaceDict_q}(\kernel_q(\cdot,\vecInputNodeTime))||_{\spaceRKHS_q}^2 &= \langle P_{\spaceDict_q}(\kernel_q(\cdot,\vecInputNodeTime)),\kernel_q(\cdot,\vecInputNodeTime) \rangle_{\spaceRKHS_q} \\
	&= \sum_{\idxDict=1}^{\noDict} \alpha_{\idxNode,\idxDict}^{(q)} \kernel_q(\vecDictIdx,\vecInputNodeTime).
\end{align*}
Thus, the denominators in \eqref{eq:functionalProjection} can be computed by
\begin{align*}
	\sum_{q\in\setKernels}||P_{\spaceDict_q}(\kernel_q(\cdot,\vecInputNodeTime))||_{\spaceRKHS_q}^2 &= \sum_{q\in\setKernels} \sum_{\idxDict=1}^{\noDict} \alpha_{\idxNode,\idxDict}^{(q)} \kernel_q(\vecDictIdx,\vecInputNodeTime) \\
	&= ||\matKernel^{-1}\vecKernel(\vecInputNodeTime)||_{\matKernel}^2.
\end{align*}
By parameterizing each $\Phi_\idxNode$ in terms of $\vecWeightNode$ and assuming a fixed dictionary $\setDict_q$ for each kernel $\kernel_q$ we arrive at the update equation presented in \eqref{eq:localApsm} with the projection \eqref{eq:projection_slab}.
}

\section{Proof of lemma \ref{lem:modConsensusMat}} \label{sec:proof_modConsensusNorm}
Let us consider the squared $\matKernelFull$-norm of $\widehat{\matConsensus}$:
\begin{equation*}
	||\widehat{\matConsensus}||_{\matKernelFull}^2 := \max_{\vecc{x}\neq\vecc{0}}\frac{||\widehat{\matConsensus}\vecc{x}||_{\matKernelFull}^2}{||\vecc{x}||_{\matKernelFull}^2}
	= \max_{\vecc{x}\neq\vecc{0}}\frac{\vecc{x}^\T\widehat{\matConsensus}^\T\matKernelFull\widehat{\matConsensus}\vecc{x}}{\vecc{x}^\T \matKernelFull \vecc{x}}
\end{equation*}
We assume $\matKernelFull$ to be non-singular. Hence, $\matKernelFull^{\nicefrac{-1}{2}} $ exists and we can insert $\widehat{\matConsensus}=\matKernelFull^{\nicefrac{-1}{2}}\matConsensus\matKernelFull^{\nicefrac{1}{2}}$:
\begin{align*}
	||\widehat{\matConsensus}||_{\matKernelFull}^2 &= \max_{\vecc{x}\neq\vecc{0}}\frac{ \vecc{x}^\T\matKernelFull^{\nicefrac{1}{2}}\matConsensus^\T\matKernelFull^{\nicefrac{-1}{2}}\matKernelFull\matKernelFull^{\nicefrac{-1}{2}}\matConsensus\matKernelFull^{\nicefrac{1}{2}}\vecc{x} }{\vecc{x}^\T \matKernelFull \vecc{x}}\\
	&= \max_{\vecc{y}\neq\vecc{0}}\frac{ \vecc{y}^\T\matConsensus^\T\matConsensus\vecc{y} }{\vecc{y}^\T \vecc{y}} \text{ with } \vecc{y}=\matKernelFull^{\nicefrac{1}{2}}\vecc{x}\\
\end{align*}
By definition of the consensus matrix $\matConsensus$ it follows that $||\widehat{\matConsensus}||_{\matKernelFull}^2=1$.
We now show that the modified consensus matrix $\widehat{\matConsensus}$ is identical to $\matConsensus$. Assume that $\matConsensus$ is compatible to the graph $\setGraph$ of any connected, undirected network via the matrix $\matEdges\in\R^{\noNodes\times\noNodes}$. Then, it holds that $\matConsensus=\matEdges\otimes\matEye_{\noDict\noKernels}$. By examining the definition of $\widehat{\matConsensus}$ we find 
\begin{align*}
	&\widehat{\matConsensus} = \matKernelFull^{-1/2}\matConsensus\matKernelFull^{1/2}=\matKernelFull^{-1/2}(\matEdges\otimes\matEye_{\noDict\noKernels})\matKernelFull^{1/2}\\
	&= \left[\begin{array}{ccc}
		\matKernel^{-1/2} &  & \vecc{0} \\
		& \ddots &  \\
		\vecc{0} &  & \matKernel^{-1/2}
		\end{array}\right]\\
	&\times
		\left[\begin{array}{ccc}
		g_{11}\matEye_{\noDict\noKernels} & \dots & g_{1\noNodes}\matEye_{\noDict\noKernels} \\
		\vdots & \ddots & \vdots \\
		g_{J1}\matEye_{\noDict\noKernels} & \dots & g_{\noNodes\noNodes}\matEye_{\noDict\noKernels}
		\end{array}\right] 
		\left[\begin{array}{ccc}
		\matKernel^{1/2} &  & \vecc{0} \\
		& \ddots &  \\
		\vecc{0} &  & \matKernel^{1/2}
		\end{array}\right]\\
	&= \left[\begin{array}{ccc}
		g_{11}\matKernel^{-1/2}\matEye_{\noDict\noKernels}\matKernel^{1/2} & \dots & g_{1\noNodes}\matKernel^{-1/2}\matEye_{\noDict\noKernels}\matKernel^{1/2} \\
		\vdots & \ddots & \vdots \\
		g_{J1}\matKernel^{-1/2}\matEye_{\noDict\noKernels}\matKernel^{1/2} & \dots & g_{\noNodes\noNodes}\matKernel^{-1/2}\matEye_{\noDict\noKernels}\matKernel^{1/2}
		\end{array}\right]\\
	&= \matEdges\otimes\matEye_{\noDict\noKernels} = \matConsensus
\end{align*}
Thus, matrices $\widehat{\matConsensus}$ and $\matConsensus$ are equivalent to each other. 

\section{Proof of Theorem \ref{theorem:analysis}} \label{sec:proof_analysis}

\subsection{Proof of Theorem \ref{theorem:analysis}.\ref{item:convergence}}
We mimic the proof of \cite[Theorem 2.3]{Cavalcante2013} to show the convergence of $(\vecc{z}_\idxIter)_{\idxIter\in\N}$. From Theorem \ref{theorem:analysis}.\ref{item:monotone} we know that the sequence $(||\vecc{z}_\idxIter-\vecc{z}^\star||^2_{\matKernelFull})_{\idxIter\in\N}$ converges for every $\vecc{z}^\star = [(\vecWeightOpt)^\T,\ldots,(\vecWeightOpt)^\T]^\T$ where $\vecWeightOpt\in\setSolutionOpt$. Thus, the sequence $(\vecc{z}_\idxIter)_{\idxIter\in\N}$ is bounded and every subsequence of $(\vecc{z}_\idxIter)_{\idxIter\in\N}$ has an accumulation point. Then, according to the Bolzano-Weierstrass Theorem the bounded real sequence $(\vecc{z}_\idxIter)_{\idxIter\in\N}$ has a convergent subsequence $(\vecc{z}_{\idxIter_l})_{\idxIter_l\in\N}$. Let $\widehat{\vecc{z}}$ be the unique accumulation point of $(\vecc{z}_{\idxIter_l})_{\idxIter_l\in\N}$. With $\lim_{\idxIter\to\infty} (\matEye_{\noDict\noKernels\noNodes}-\mat{B}\mat{B}^\T)\vecc{z}_{\idxIter} = \vecc{0}$ it follows that 
$$\lim_{\idxIter\to\infty} (\matEye_{\noDict\noKernels\noNodes}-\mat{B}\mat{B}^\T)\vecc{z}_{\idxIter_l} = (\matEye_{\noDict\noKernels\noNodes}-\mat{B}\mat{B}^\T) \widehat{\vecc{z}} = \vecc{0}.$$ 
Hence, $\widehat{\vecc{z}}$ lies in the consensus subspace $\mathcal{C}$. To show that this point is a unique accumulation point suppose the contrary, i.e., $\widehat{\vecc{z}}=[\widehat{\vecWeight}^\T,\ldots,\widehat{\vecWeight}^\T]^\T\in\mathcal{C}$ and $\tilde{\vecc{z}}=[\tilde{\vecWeight}^\T,\ldots,\tilde{\vecWeight}^\T]^\T\in\mathcal{C}$ are two different accumulation points. For every $\vecc{z}^\star$ the sequence $(||\vecc{z}_\idxIter-\vecc{z}^\star||_{\matKernelFull}^2)_{\idxIter\in\N}$ converges and hence it follows that
\begin{align*}
	0 &= ||\widehat{\vecc{z}}-\vecc{z}^\star||_{\matKernelFull}^2 - ||\tilde{\vecc{z}}-\vecc{z}^\star||_{\matKernelFull}^2 \\
	  &= ||\widehat{\vecc{z}}||_{\matKernelFull}^2 - ||\tilde{\vecc{z}}||_{\matKernelFull}^2 - 2(\widehat{\vecc{z}}-\tilde{\vecc{z}})^\T \matKernelFull \vecc{z}^\star\\
	  &= ||\widehat{\vecc{z}}||_{\matKernelFull}^2 - ||\tilde{\vecc{z}}||_{\matKernelFull}^2 - 2\noNodes(\widehat{\vecWeight}-\tilde{\vecWeight})^\T \matKernel \vecWeightOpt.
\end{align*}
It thus holds that $\vecWeightOpt\in H:=\big\{ \vecWeight \, | \, 2\noNodes (\widehat{\vecWeight}-\tilde{\vecWeight})\matKernel \vecWeight = ||\widehat{\vecc{z}}||_{\matKernelFull}^2 - ||\tilde{\vecc{z}}||_{\matKernelFull}^2 \big\}$ where $\widehat{\vecWeight}-\tilde{\vecWeight}\neq 0\, (\Leftrightarrow \widehat{\vecc{z}}\neq\tilde{\vecc{z}} )$. Since we assume that $\vecWeightOpt\in\setSolutionOpt$ this implies that $\setSolutionOpt$ is a subset of the hyperplane~$H$. This contradicts the assumption of a nonempty interior of $\setSolutionOpt$. Hence, the bounded sequence $(\vecc{z}_\idxIter)_{\idxIter\in\N}$ has a unique accumulation point, and so it converges.

\subsection{Proof of Theorem \ref{theorem:analysis}.\ref{item:characterization}}
In this proof we mimic the proof of \cite[Theorem 2(d)]{Yamada2004}, \cite[Theorem 3.1.4]{Slavakis2006} and \cite[Theorem 2(e)]{Cavalcante2011} to characterize the limit point $\widehat{\vecWeight}$ of the sequence $(\vecWeightNodeTime)_{\idxIter\in\N},\forall\idxNode\in\setNodes$. Furthermore, we need Claim 2 of \cite{Yamada2004} which is proven for any real Hilbert space and thus, also holds for the $\matKernel$-metric Euclidean space considered here. 
\begin{fact}[Claim 2 in \cite{Yamada2004}] \label{fact1}
Let $C\subset\R^{\noDict\noKernels}$ be a nonempty closed convex set. Suppose that $\rho>0$ and $\tilde{\vecc{u}}$ satisfies $\{ \vecc{v}\in\R^{\noDict\noKernels} \,|\, ||\vecc{v}-\tilde{\vecc{u}}||_{\matKernel} \leq \rho \}\subset C$. Assume $\vecWeight\in\R^{\noDict\noKernels}/C$ and $t\in(0,1)$ such that $\vecc{u}_t:= t\vecWeight+(1-t)\tilde{\vecc{u}}\notin C$. Then $\dis_{\matKernel}(\vecWeight,C)>\rho \frac{1-t}{t} = \rho \frac{||\vecc{u}_t-\vecWeight||_{\matKernel}}{||\vecc{u}_t-\tilde{\vecc{u}}||_{\matKernel}}>0$ with $\dis_{\matKernel}(\vecc{w},C):=||\vecc{w}-P^{\mat{K}}_{C}(\vecc{w})||_{\matKernel}$.
\end{fact}

Assume the contrary of our statement, i.e., $\widehat{\vecWeight} \not\in \overline{\lim \inf_{\idxIter\to\infty}\, \setSolutionTime}$. Denote by $\tilde{\vecc{u}}$ an interior point of $\setSolutionOpt$. Therefore, there exists $\rho>0$ such that $\{ \vecc{v}\in\R^{\noDict\noKernels} \,|\, ||\vecc{v}-\tilde{\vecc{u}}||_{\matKernel} \leq \rho \}\subset\setSolutionOpt$. Furthermore, there exists $t\in(0,1)$ such that $\vecc{u}_t:= t\widehat{\vecWeight} + (1-t) \tilde{\vecc{u}}\notin \setSolution \supset \lim \inf_{\idxIter\to\infty} \setSolutionTime$. Since $\lim_{\idxIter\to\infty} \vecWeightNodeTime = \widehat{\vecWeight}\, (\forall\idxNode\in\setNodes)$ there exists $N_1\in\N$ such that $||\vecWeightNodeTime-\widehat{\vecWeight}||_{\matKernel} \leq \rho \frac{1-t}{2t}, \forall \idxIter\geq N_1,\forall \idxNode\in\setNodes$. Then, by $\vecc{u}_t \notin  \lim \inf_{\idxIter\to\infty} \setSolutionTime$ for any $L_1>N_1$ there exists $\idxIter_1 \geq L_1$ satisfying $\vecc{u}_t \notin \setSolution_{\idxIter_1} = \bigcap_{\idxNode\in\setNodes} (\lev_{\leq0} \funCost_{\idxNode,\idxIter_1})$ where $\lev_{\leq0} \funCost_{\idxNode,\idxIter_1}:=\{ \vecWeight\in\R^{\noDict\noKernels} \,|\, \funCost_{\idxNode,\idxIter_1}(\vecWeight) \leq 0 \}$.
It follows that there exists a node $i\in\setNodes$ such that $\vecc{u}_t\notin \lev_{\leq 0} \funCost_{i,\idxIter_1}$. By $\setSolution\subset\setSolutionTime\subset\lev_{\leq0} \funCost_{i,\idxIter_1}$ and Fact \ref{fact1} for node $i$ it holds that
\begin{align*}
	\dis_{\matKernel}(\vecWeight_{i,\idxIter_1},\lev_{\leq 0} \funCost_{i,\idxIter_1}) &\geq \dis_{\matKernel}(\widehat{\vecWeight},\lev_{\leq 0} \funCost_{i,\idxIter_1})\\ &- ||\vecWeight_{i,\idxIter_1}-\widehat{\vecWeight}||_{\matKernel} \\
	& \geq \rho \frac{1-t}{t} -\frac{\rho}{2} \frac{1-t}{t}\\ 
	&=  \frac{\rho}{2} \frac{1-t}{t} =: \epsilon >0
\end{align*}
Thus, it follows that $\sum_{\idxNode\in\setNodes} \dis_{\matKernel}(\vecWeight_{\idxNode,\idxIter_1},\lev_{\leq 0} \funCost_{\idxNode,\idxIter_1}) \geq \epsilon$. By the triangle inequality we have
\begin{align*}
	||\tilde{\vecc{u}} - \vecWeight_{\idxNode,\idxIter_1}||_{\matKernel} &\leq ||\tilde{\vecc{u}} - \widehat{\vecWeight}||_{\matKernel}+||\vecWeight_{\idxNode,\idxIter_1} - \widehat{\vecWeight}||_{\matKernel}\\ 
	&\leq ||\tilde{\vecc{u}} - \widehat{\vecWeight}||_{\matKernel} + \frac{\rho}{2} \frac{1-t}{t}\quad (\idxNode\in\setNodes)
\end{align*}
so that
\begin{equation*}
	\sum_{\idxNode\in\setNodes} ||\tilde{\vecc{u}} - \vecWeight_{\idxNode,\idxIter_1}||_{\matKernel} \leq \noNodes ||\tilde{\vecc{u}} - \widehat{\vecWeight}||_{\matKernel}+ \noNodes \frac{\rho}{2} \frac{1-t}{t} =: \eta > 0.
\end{equation*}
Given a fixed $L_2>\idxIter_1$, we can find a $\idxIter_2 \geq L_2$ such that $\sum_{\idxNode\in\setNodes} \dis_{\matKernel}(\vecWeight_{\idxNode,\idxIter_2},\lev_{\leq 0} \funCost_{\idxNode,\idxIter_2}) \geq \epsilon$ and $\sum_{\idxNode\in\setNodes} ||\tilde{\vecc{u}} - \vecWeight_{\idxNode,\idxIter_2}||_{\matKernel} \leq \eta$. Thus, we can construct a subsequence $\{\idxIter_l\}_{l=1}^\infty$ satisfying
\begin{equation*}
	\sum_{\idxNode\in\setNodes} \dis_{\matKernel}(\vecWeight_{\idxNode,\idxIter_l},\lev_{\leq 0} \funCost_{\idxNode,\idxIter_l}) \geq \epsilon \text{ and }
	\sum_{\idxNode\in\setNodes} ||\tilde{\vecc{u}} - \vecWeight_{\idxNode,\idxIter_l}||_{\matKernel} \leq \eta.
\end{equation*}
With the assumptions of the theorem there exists a $\zeta>0$ such that $\sum_{\idxNode\in\setNodes} \funCost_{\idxNode,\idxIter_l}(\vecWeight_{\idxNode,\idxIter_l}) \geq \zeta$ for every $l\geq 1$. However, this contradicts $\lim_{\idxIter\to\infty} \funCostNodeTime(\vecWeightNodeTime)=0, \forall\idxNode\in\setNodes$ from Theorem~\ref{theorem:analysis}.\ref{item:asymptotic_min}. Thus, it follows that $\widehat{\vecWeight} \in \overline{\lim \inf_{\idxIter\to\infty} \setSolutionTime}$ and the proof is complete.

\appendices



\bibliographystyle{IEEEtran}
%


\begin{IEEEbiography}[{\includegraphics[width=1in,height=1.25in,clip,keepaspectratio]{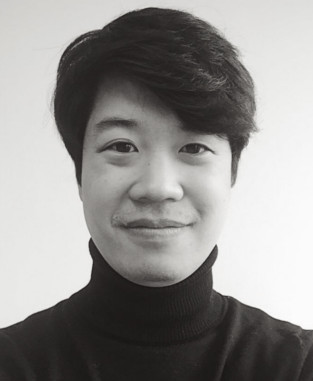}}]{Ban-Sok Shin}
(S’13) received the Dipl.-Ing. (M.Sc.) degree from University of Bremen in 2013. Since then, he has been a research assistant at the Department of Communications Engineering of the University of Bremen where he is currently working towards his Ph.D. degree. His research interests include distributed inference/estimation, adaptive signal processing, machine learning and their application to sensor networks.
\end{IEEEbiography}

\begin{IEEEbiography}[{\includegraphics[width=1in,height=1.25in,clip,keepaspectratio]{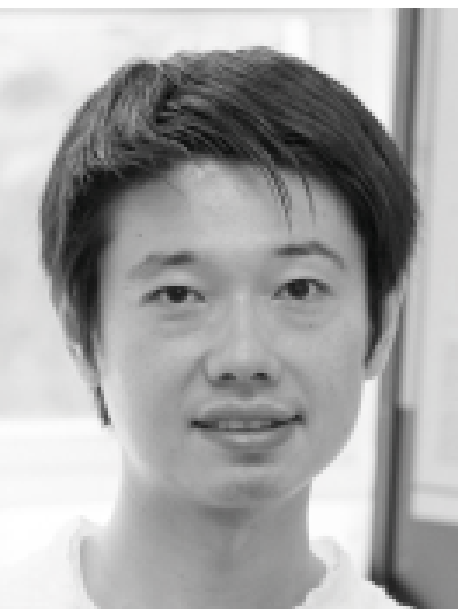}}]{Masahiro Yukawa}
(S’05-M’06) received the B.E.,
M.E., and Ph.D. degrees from Tokyo Institute of
Technology in 2002, 2004, and 2006, respectively.
He studied as Visiting Researcher/Professor with the
University of York, U.K. (Oct.2006-March 2007),
with the Technical University of Munich, Germany (July 2008-November 2007),
and with the Technical University of Berlin, Germany (April 2016-February 2017).
He worked with RIKEN, Japan, as Special Postdoctoral Researcher (2007-2010), and
with Niigata University, Japan, as Associate Professor (2010-2013).
He is currently Associate Professor with the Department of Electronics
and Electrical Engineering, Keio University, Japan.
Since July 2017, he has been Visiting Scientist at AIP Center, RIKEN, Japan.
He has been Associate Editor for the IEEE \textsc{Transactions on Signal Processing} (since 2015),
Multidimensional Systems and Signal Processing (2012-2016), and the IEICE
Transactions on Fundamentals of Electronics, Communications and Computer
Sciences (2009-2013). His research interests include mathematical adaptive
signal processing, convex/sparse optimization, and machine learning.

Dr. Yukawa was a recipient of the Research Fellowship of the Japan Society
for the Promotion of Science (JSPS) from April 2005 to March 2007. He
received the Excellent Paper Award and the Young Researcher Award from
the IEICE in 2006 and in 2010, respectively, the Yasujiro Niwa Outstanding
Paper Award in 2007, the Ericsson Young Scientist Award in 2009, the
TELECOM System Technology Award in 2014, the Young Scientists’ Prize,
the Commendation for Science and Technology by the Minister of Education,
Culture, Sports, Science and Technology in 2014, the KDDI Foundation
Research Award in 2015, and the FFIT Academic Award in 2016. He is
a member of the IEICE.

\end{IEEEbiography}

\begin{IEEEbiography}[{\includegraphics[width=1in,height=1.25in,clip,keepaspectratio]{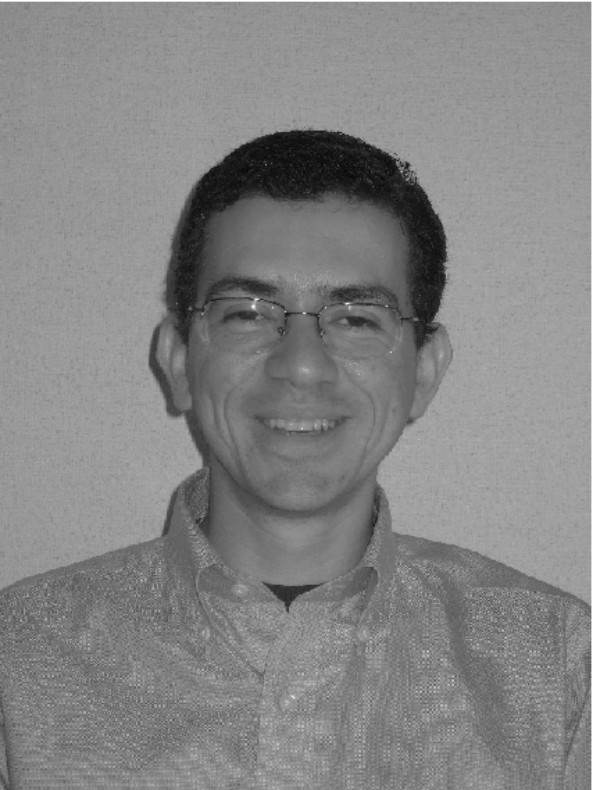}}]{Renato Lu\'is Garrido Cavalcante}
(M’09) received the electronics engineering degree from the Instituto Tecnol\'ogico de Aeron\'autica (ITA), Brazil, in 2002, and the M.E. and Ph.D. degrees in Communications and Integrated Systems from the Tokyo Institute of
Technology, Japan, in 2006 and 2008, respectively. From April 2003 to April 2008, he was a recipient of the Japanese Government (MEXT) Scholarship. \par
He is currently a Research Fellow with the Fraunhofer Institute for Telecommunications, Heinrich Hertz Institute, Berlin, Germany, and a lecturer at the Technical University of Berlin. Previously, he held appointments as a Research Fellow with the University of Southampton, Southampton, U.K., and as a Research Associate with the University of Edinburgh, Edinburgh, U.K. \par
Dr. Cavalcante received the Excellent Paper Award from the IEICE in 2006 and the IEEE Signal Processing Society (Japan Chapter) Student Paper Award in 2008. He also co-authored a study that received a best student paper award at the 13th IEEE International Workshop on Signal Processing Advances in Wireless Communications (SPAWC) in 2012. His current interests are in signal processing for distributed systems, multiagent systems, convex analysis, machine learning, and wireless communications.
\end{IEEEbiography}
%
%

\begin{IEEEbiography}[{\includegraphics[width=1in,height=1.25in,clip,keepaspectratio]{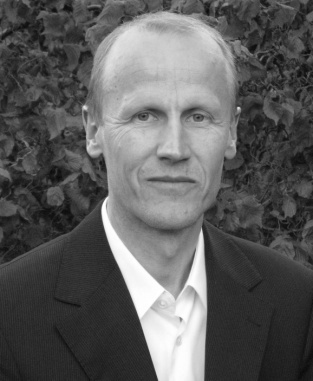}}]{Armin Dekorsy}
received the B.Sc. degree from Fachhochschule Konstanz, Konstanz, Germany, the M.Sc. degree from the University of Paderborn, Paderborn, Germany, and the Ph.D. degree from the University of Bremen, Bremen, Germany, all in communications engineering.

From 2000 to 2007, he was a Research Engineer at Deutsche Telekom AG and a Distinguished Member of Technical Staff at Bell Labs Europe, Lucent Technologies. In 2007, he joined Qualcomm GmbH as a European Research Coordinator, conducting Qualcomms internal and external European research activities. He is currently
the Head of the Department of Communications Engineering, University of Bremen. He has authored or coauthored over 150 journal and conference publications and is the holder of over 17 patents in the area of wireless communications. He has long-term expertise in the research of wireless communication systems, baseband algorithms, and signal processing.

Prof. Dekorsy is a senior member of the IEEE Communications and Signal Processing Society and the VDE/ITG expert committee “Information and System Theory.”
\end{IEEEbiography}


\end{document}